\documentclass[letterpaper]{article} % DO NOT CHANGE THIS
\usepackage{aaai25}  % DO NOT CHANGE THIS
\usepackage{times}  % DO NOT CHANGE THIS
\usepackage{helvet}  % DO NOT CHANGE THIS
\usepackage{courier}  % DO NOT CHANGE THIS
\usepackage[hyphens]{url}  % DO NOT CHANGE THIS
\usepackage{graphicx} % DO NOT CHANGE THIS
\usepackage{natbib}  % DO NOT CHANGE THIS AND DO NOT ADD ANY OPTIONS TO IT
\usepackage{caption} % DO NOT CHANGE THIS AND DO NOT ADD ANY OPTIONS TO IT
\usepackage{threeparttable}
\usepackage{multirow}

\usepackage{pgfplots}
\usepackage{bm}
\usepackage{enumitem}

\usepackage{algorithm}
\usepackage{algorithmic}

\usepackage{amsmath}
\usepackage{amssymb}
\usepackage{mathtools}
\usepackage{amsthm}
\usepackage{pifont}
\usepackage{booktabs}

\theoremstyle{plain}
\newtheorem{theorem}{Theorem}

\newtheorem{lemma}[theorem]{Lemma}
\newtheorem{corollary}[theorem]{Corollary}

\theoremstyle{definition}
\newtheorem{definition}[theorem]{Definition}
\newtheorem{assumption}[theorem]{Assumption}
\newtheorem{remark}[theorem]{Remark}

\usepackage{newfloat}
\usepackage{listings}
\DeclareCaptionStyle{ruled}{labelfont=normalfont,labelsep=colon,strut=off} % DO NOT CHANGE THIS
\floatstyle{ruled}
\newfloat{listing}{tb}{lst}{}
\floatname{listing}{Listing}
\title{UFID: A Unified Framework for Black-box Input-level Backdoor Detection on Diffusion Models}
\author{
    Zihan Guan\equalcontrib \textsuperscript{\rm 1,\rm 2}, Mengxuan Hu\equalcontrib \textsuperscript{\rm 3}, Sheng Li$^\dagger\textsuperscript{\rm 3}$  , Anil Kumar Vullikanti\thanks{Co-corresponding Author}\textsuperscript{\rm 1,\rm 2}
}
\affiliations{
    \textsuperscript{\rm 1}Department of Computer Science, University of Virginia \\
    \textsuperscript{\rm 2}Biocomplexity Institute, University of Virginia \\
    \textsuperscript{\rm 3}School of Data Science, University of Virginia \\
    \{bxv6gs, qtq7su, shengli, vsakumar\}@virginia.edu
}

\usepackage{bibentry}
\begin{document}

\maketitle

\begin{abstract}
Diffusion models are vulnerable to backdoor attacks, where malicious attackers inject backdoors by poisoning certain training samples during the training stage. This poses a significant threat to real-world applications in the Model-as-a-Service (MaaS) scenario, where users query diffusion models through APIs or directly download them from the internet. To mitigate the threat of backdoor attacks under MaaS, \textit{black-box input-level backdoor detection} has drawn recent interest, where defenders aim to build a firewall that filters out backdoor samples in the inference stage, with access only to input queries and the generated results from diffusion models. Despite some preliminary explorations on the traditional classification tasks, these methods cannot be directly applied to the generative tasks due to two major challenges: (1) more diverse failures and (2) a multi-modality attack surface. In this paper, we propose a black-box input-level backdoor detection framework on diffusion models, called UFID. Our defense is motivated by an insightful causal analysis: Backdoor attacks serve as the confounder, introducing a spurious path from input to target images, which remains consistent even when we perturb the input samples with Gaussian noise. We further validate the intuition with theoretical analysis. Extensive experiments across different datasets on both conditional and unconditional diffusion models show that our method achieves superb performance on detection effectiveness and run-time efficiency.
\end{abstract}

\section{Introduction}
Diffusion models~\citep{ho2020denoising, song2020improved, song2020score, song2020denoising} have emerged as the new state-of-the-art family of generative models due to their superior performance~\citep{dhariwal2021diffusion} and wide applications across a variety of domains, ranging from computer vision~\citep{ baranchuk2021label, brempong2022denoising}, natural language processing~\citep{austin2021structured, hoogeboom2021argmax, li2022diffusion}, and robust machine learning~\citep{blau2022threat, carlini2022certified}. 
% Despite their success, the training of diffusion models consumes a tremendous amount of time and computational resources. Consequently, it is common practice to directly utilize third-party models via an API or directly download them from the internet. The setting is referred to as Model-as-a-Service (MaaS).
Despite their success, training diffusion models requires significant time and computational resources. Consequently, it is common practice to utilize third-party models via an API or to download them directly from the internet. This approach is known as Model-as-a-Service (MaaS).

However, it has recently been found that diffusion models are vulnerable to backdoor attacks~\cite{gu2024responsible, Huang_Juefei-Xu_Guo_Zhang_Wu_Hu_Li_Pu_Liu_2024, chen2023trojdiff, Chou_2023_CVPR, chou2023villandiffusion, struppek2022rickrolling, guan2023Attacking}, where malicious attackers poison certain training samples with a predefined trigger pattern in the training stage. 
% Afterward, the diffusion models' behavior will be adversarially manipulated whenever the trigger pattern is present in the input query but normal when the input query is clean. This vulnerability not only poses a serious threat to the downstream users, for example, generating erotic images to children who query it, but also the companies or artise who owns extensive copyright,  for example, manipulate the model to generate image without the copyright.
Consequently, the behavior of diffusion models can be adversarially manipulated whenever the trigger pattern appears in the input query, while it remains normal with clean input queries. This vulnerability poses a serious threat to real-world applications in the MaaS setting, e.g., the online third-party diffusion models may have been backdoored to generate inappropriate images for children or to generate images that bypass copyright restrictions when a specific trigger appears in the query ~\cite{wang2024stronger}.
More seriously, the traditional training-phase defense methods~\cite{li2021anti, huang2022backdoor,gao2023backdoor} cannot be deployed in the MaaS setting due to the defenders' inaccessibility to the training pipeline and training data.

To mitigate the above threat, \textit{black-box input-level backdoor detection} has recently drawn great interest. In this scenario, \textit{input-level} indicates that defenders aim to build a firewall-style detector in the inference stage to filter out and reject backdoored inputs while allowing clean inputs to generate predictions. \textit{Black-box} means that defenders only have access to user queries and the generated results from the deployed models, without any prior information (e.g., model weights, architectures) assumed by the previous works~\cite{ma2022beatrix,qiu2021deepsweep,gao2021design,gao2019strip}.

% \begin{table}[!t]
%     \centering
%     \caption{Comparison of the problem settings.}\label{tab:comparison}
%     \vspace{-2mm}
%     \resizebox{\columnwidth}{!}{
%     % \vspace{-4mm}
%     % \small
%     % \setlength{\tabcolsep}{1mm}
%     \begin{tabular}{c|cccc}
%     \toprule
%          Task & Method & Black-box &  Diverse Failures & Multi-Modality Attack Surface \\
%          \midrule
%          \multirow{5}{*}{Classification} & Beatrix~\cite{ma2022beatrix} & \emptycirc & \emptycirc & \emptycirc \\
%          &STRIP~\cite{gao2019strip} & \emptycirc & \emptycirc & \emptycirc \\
%          &SCALE-UP~\cite{guo2023scaleup} & \fullcirc & \emptycirc & \emptycirc\\
%          &TeCO~\cite{liu2023detecting} & \fullcirc & \emptycirc & \emptycirc \\
%          \midrule
%          \multirow{3}{*}{Generative} &TERD~\cite{mo2024terd} & \emptycirc & \halfcirc & \emptycirc \\
%          &Shield~\cite{Wang2024T2IShield} & \emptycirc & \fullcirc & \halfcirc\\
%          % \midrule
%           &\textbf{UFID (Ours)} & \fullcirc & \fullcirc & \fullcirc\\
%     \bottomrule
%     \end{tabular}
%     }
%     \vspace{-5mm}
% \end{table}

Prior methods for backdoor detection in image classification, e.g.,~\cite{guo2023scaleup,liu2023detecting, hu2024bbcal} cannot be adopted for generative tasks due to two major challenges: \ding{182} \textbf{More Diverse Failures}: Unlike merely generating a fixed target image (e.g., a Hello-Kitty image), backdoored diffusion models can be manipulated to produce a specific class of images (e.g., cat images), or even images with a specified abstract concept (e.g., erotic images). This implies that the target images are not necessarily unique but can vary as long as they belong to the designated target class. This variability substantially complicates the detection in the generative task. \ding{183} \textbf{Multi-Modality Attack Surface}: Unlike traditional image classifiers, which involve a single modality, diffusion models (e.g., Stable Diffusion) are capable of supporting multiple modalities. This diversity necessitates a unified framework for backdoor detection in diffusion models. An overall comparison of the problem setting is in the Appendix.

To address the above challenges, our intuition is motivated by a causal analysis: backdoor attacks serve as the confounder, introducing a spurious path from input to target images. The spurious path embedded in the diffusion model remains consistent even when we perturb input samples with Gaussian noise. 
Therefore, the backdoor generations remain consistent after minor perturbation, while the clean generations alter significantly even with a small perturbation. 
% We further provide a theoretical basis for this causal analysis, and show that the distance between the backdoor generations and clean generations after perturbation can be lower bounded (Corollary~\ref{corollary:large_quantify}).
We further validate the causal analysis rigorously, showing that after perturbation, the difference between the diversity of clean generations and the diversity of backdoor generations is lower bounded (Corollary~\ref{corollary:large_quantify}). 
Driven by the analysis, we develop a \textbf{U}nified \textbf{F}ramework for black-box \textbf{I}nput-level backdoor \textbf{D}etection (UFID) on diffusion models. Specifically, UFID examines each input sample by calculating its \textit{graph density score}, which measures the similarity within the generated batch after perturbing the given input sample. The higher the graph density score, the more likely the input sample is backdoored. Compared to the existing method TERD~\cite{mo2024terd}, UFID is designed for a black-box setting, requiring no access to model weights or structures. Moreover, the performance gap between UFID and TERD is also satisfactory, with a maximum difference of 8\% in precision, 7\% in Recall, and 1\% in AUC. UFID is also generalized to the scenario of conditional diffusion models, where the input can be of various modalities. To strengthen UFID's resistance to diversity-intensive backdoor attacks, we also design strategies to integrate supplementary correspondence information into the UFID framework. In contrast, Shield~\cite{Wang2024T2IShield} operates in a white-box setting but achieves similar performance as UFID.

% gradually perturb the inputs with different random noises; for conditional diffusion models, we perturb the text inputs by augmenting them with additional public textual prompts. Based on our previous analysis, a backdoored input tends to produce similar images regardless of perturbations, while a clean sample generates multiple diversified images. This different prediction behavior in image generation allows us to use the similarity of the generated images as a criterion for detecting backdoor samples. 

% Extensive experiments across different datasets on both unconditional and conditional diffusion models demonstrate that our method achieves superb performance on detection effectiveness and run-time efficiency.

% To simply put, we leverage this property to detect backdoor samples by gradually adding noise on each input in backdoored unconditional diffusion models and augmenting the text inputs with additional phrases in backdoored conditional diffusion models. While a backdoored input tends to produce similar images regardless of the noise or masking, a clean sample will generate multiple different images. This contrast in image generation allows us to use the similarity of the generated images as a basis for detecting backdoor samples. Extensive experiments across different datasets on both conditional and unconditional diffusion models demonstrate that our method achieves superb performance on detection effectiveness and run-time efficiency.

To sum up, our contributions in the work include: (1) \textbf{First Unified Black-box Backdoor Detection Framework for Diffusion Models}. To the best of our knowledge, our work is the first unified black-box framework for detecting input-level backdoor samples in diffusion models; (2) \textbf{Novel Causality Analysis}. We apply causality analysis for analyzing backdoor attacks on generative tasks; Besides, theoretical analysis also sheds light on our intuitions and validates the effectiveness of our method; (3) \textbf{Promising Performance}. Extensive results show that our detection method achieves an average of nearly 100\% AUC on the unconditional models, and 90\% AUC on the conditional models with an acceptable inference overhead.\footnote{An extended version of this work with appendices and further
details is available at: {https://arxiv.org/abs/2404.01101}}

\section{Related Works}
% \paragraph{Diffusion Models.}
% Diffusion models have become a new state-of-the-art family of generative models in image synthesis by showing strong sample quality and diversity in image synthesis~\cite{ho2020denoising,song2020denoising}. Since then, diffusion models have been widely applied to different types of tasks, such as image generation~\cite{song2019generative,song2020score}, image super-resolution~\cite{saharia2022image,batzolis2021conditional}, and image editing~\cite{avrahami2022blended,choi2021ilvr}. Moreover, researchers also found that the representations learned by diffusion models can also be used in other discriminative tasks such as segmentation~\cite{graikos2022diffusion} and anomaly detection~\cite{pinaya2022fast}.

\noindent
\textbf{Backdoor Attacks and Defenses on Diffusion Models.} Recently, a lot of works have investigated the security vulnerabilities of diffusion models by launching backdoor attacks on diffusion models. From a high-level idea, malicious backdoor attackers aim to inject a special behavior in the diffusion process such that, once the predefined trigger pattern appears on the input, the special behaviors will be activated. To achieve this goal, ~\cite{chen2023trojdiff} proposed to add an additional backdoor injection task on the training stage and maliciously alter the sampling procedure with a correction term. ~\cite{Chou_2023_CVPR} proposed a novel attacking strategy by only modifying the training loss function. ~\cite{an2023elijah} focused on launching attacks to text-to-image tasks, by injecting backdoors into the pre-trained text encoder. ~\cite{Huang_Juefei-Xu_Guo_Zhang_Wu_Hu_Li_Pu_Liu_2024} proposed to apply personalization techniques to efficiently inject a malicious concept into the diffusion models. ~\cite{chou2023villandiffusion} proposed a unified framework that covers all the popular schemes of diffusion models, including conditional and unconditional diffusion models. ~\cite{zhai2023text} proposed a novel backdoor attack that can generate backdoor images as diversified as clean images. Backdoor defenses on diffusion models are highly under-explored. To the best of our knowledge, only four papers~\cite{an2023elijah, mo2024terd, sui2024disdet, Wang2024T2IShield} investigated backdoor defenses on diffusion models. However, Elijah~\cite{an2023elijah} aims to detect whether a given model is backdoored, while our tasks focus on filtering backdoor samples for diffusion models in the inference stage, which are fundamentally different. TERD~\cite{mo2024terd} builds a unified framework for safeguarding diffusion models, which can handle tasks such as trigger inversion, input detection, and model detection. However, the proposed technique only works for unconditional diffusion models and requires white-box access to the weights and structures of the diffusion models. DisDet~\cite{sui2024disdet} also only focuses on unconditional diffusion models. It conducts backdoor detection by checking whether the given input follows a Gaussian distribution or not. However, the defense method can be easily circumvented by using an imperceptible trigger pattern. ~\cite{Wang2024T2IShield} proposed to detect backdoors on the text-to-image models based on a novel "assimilation phenomenon".

\section{Preliminaries}
\textbf{Diffusion Models.} Without loss of generality, diffusion models contain two parts: (1) Diffusion Process: a data distribution $q(x)$ is diffused to a target distribution $r(x)$ within $T$ timestamps. (2) Training Process: A diffusion model $\epsilon_{\theta}$ with parameter $\theta$ is trained to align with the reversed diffusion process, i.e., $p_{\theta} (x_{i-1} | x_i) = \mathcal{N}(x_{i-1}; \mu_{\theta}(x_i), \sigma_{\theta}(x_i)) = q(x_{i-1} | x_i)$. DDPM is one of the most basic diffusion models~\cite{ho2020denoising}. DDPM assumes the target distribution $r(x) = \mathcal{N}(0, I)$ and the diffusion process $q(x_i|x_{t-1}) = \mathcal{N}(x_t; \sqrt{1-\beta_i}x_{i-1}, \beta_i I)$, where the $\{\beta_i \}_{i=1}^T$ is a pre-defined variance schedule that controls the step sizes. Furthermore, let $\alpha_i = 1-\beta_i$ and $\bar{\alpha_i} = \Pi_{t=1}^i \alpha_t$. By minimizing the loss function $\| \epsilon - \epsilon_{\theta} (\sqrt{\bar{\alpha_i}} x_0 + \sqrt{1-\bar{\alpha_i}} \epsilon, i)\|^2$, the diffusion model is expected to be able to correctly predict the added noise given the input $x_i$ at time $i$. In the inference stage, DDPM generates images by sampling from the Gaussian distribution $\mathcal{N}(0, I)$ from time $i=T$ to $i=0$ with the generative process $p_{\theta}(x_{i-1} | x_i) = \mathcal{N}(x_{i-1}; \mu_{\theta}(x_i), \sigma_{\theta}(x_i))$, where $\mu_{\theta}(x_i) = \frac{1}{\alpha_i}(x_i - \frac{1-\alpha_i}{1-\bar{\alpha_i} \epsilon_{\theta}(x_i, i)})$ and $\sigma_{\theta}(x_i) = \frac{(1-\bar{\alpha}_{i-1})\beta_i}{1-\bar{\alpha}_{i}}$.

% \paragraph{DDIM.} DDIM is proposed to accelerate the sampling process of DDPM. DDIM shares the same diffusion process and loss function as the DDPM. However, DDIM leverages a different sampling strategy to accelerate the sampling procedure.

\noindent
\textbf{Backdoor Attacks on Diffusion Models.}
Different from launching backdoor attacks on the traditional models (e.g., classifiers~\cite{gu2017badnets,chen2017targeted}), which could be achieved by poisoning training dataset, injecting backdoors into the diffusion models is much more complicated. A typical backdoor attack pipeline~\cite{chen2023trojdiff, Chou_2023_CVPR, chou2023villandiffusion} on diffusion models consists of three steps: (1) the attackers first need to mathematically define the forward backdoor diffusion process, i.e., $x_0^b \rightarrow x^b_T$, where the $x_0^b$ denotes the target image and the $x_T^b$ denotes the trigger image; (2) then the attackers train the diffusion models to align with the backdoored reversed process; (3) in the inference stage, the diffusion models can be prompted to generate target images when the input contains the trigger pattern, but behave normally when the input is clean (e.g., pure Gaussian noise for the DDPM model).

\noindent
\textbf{Threat Model.}
We adopt a similar threat model as in~\citep{guo2023scaleup,gao2021design,liu2023detecting}. Specifically, the defender is assumed to only have access to user queries (e.g., prompt) and the generated results from diffusion models. The defender aims to conduct \emph{efficient} and \emph{effective} black-box backdoor detection, where efficiency requires that the detection process does not significantly impact the response time of user queries, while effectiveness requires the detection process to distinguish backdoor samples and clean samples with a high accuracy rate. The challenge of this threat model arises from three factors: \textbf{(1) More Diverse Failures}. Backdoored diffusion models can be triggered to generate specific classes of images (e.g., cat images), or even images with a specified abstract concept (e.g., erotic images), extending beyond fixed target labels in traditional classification tasks. \textbf{(2) Multi-Modality Attack Surface}. Unlike traditional image classifiers that involve only one modality, diffusion models (e.g., Stable Diffusion) can support a variety of modalities. \textbf{(3) Limited Information}. The detection method only has access to the query images and the prediction labels returned by the diffusion model.

\section{Overview of the UFID}

% extended attacks objectives , backdoor attacks on generative models have a wide range of attack objectives, including generating a static 
\subsection{Intuition: Backdoor Attacks under a Causal Lens}
\begin{figure}[!t]
  \centering
    % \vskip -0.1in
    \includegraphics[width=0.35\textwidth]{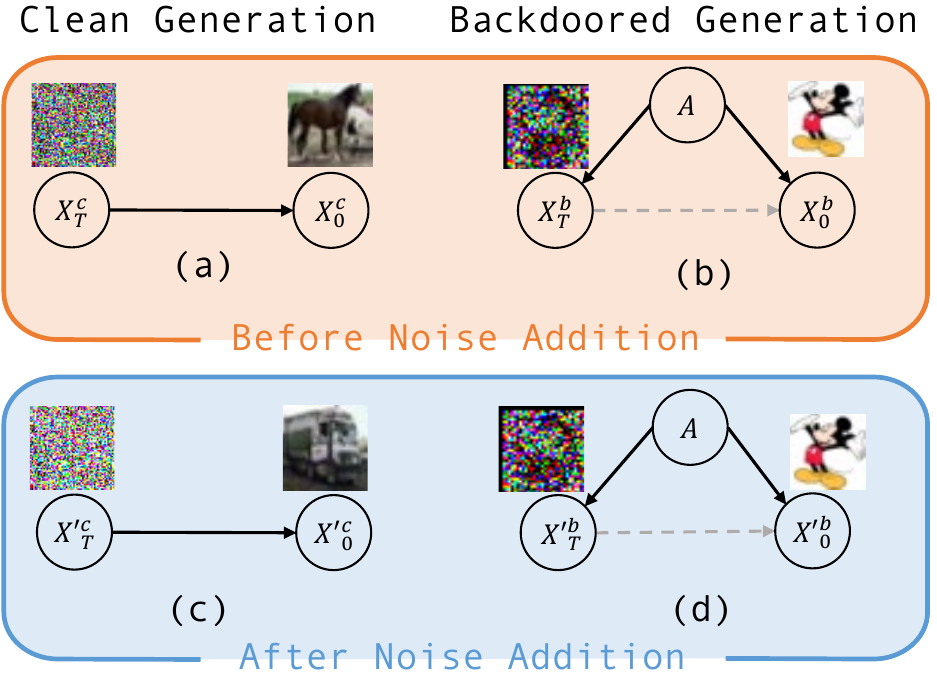}
    \caption{Causal graph of clean and backdoored generation.}
    \label{fig:causal_graph}
\end{figure}

\begin{figure*}[!t]
    \centering
    \includegraphics[width=\textwidth]{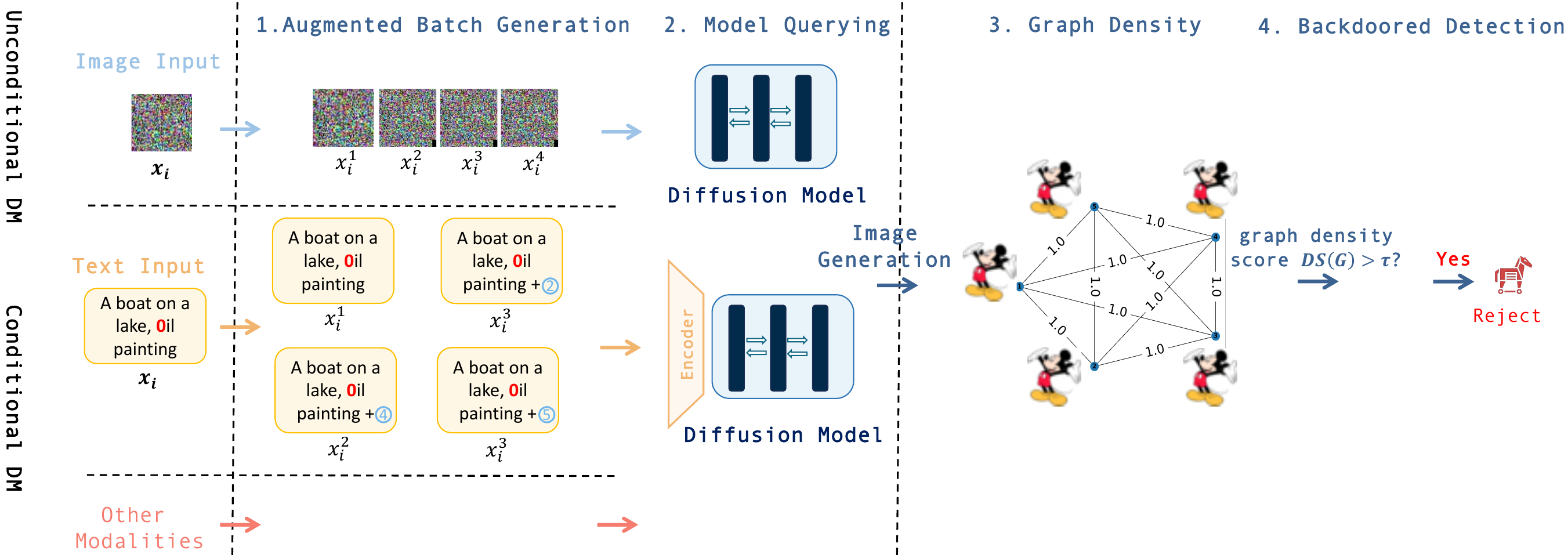}
    \caption{Pipeline of our unified framework for backdoor detection on diffusion models.}
    \label{fig:pipeline}
\end{figure*}
% To design a backdoor detection algorithm for generation models, we need to first consider a foundational question: What is the difference between clean generation and backdoored generation, and how can we leverage this difference to design a detection algorithm to distinguish them?  Hence, we propose to leverage the capability of causal inference to uncover distinct underlying mechanisms of backdoored and clean generation. Especially, the causal graph for the comparison between backdoored and clean generation process is provided in Fig.~\ref{fig:causal_graph}. A causal graph is a directed acyclic graph that indicates the causal relationships between variables, where each node in the graph represents a variable, and an edge from node $A$ towards node $B$ represents that variable $A$ serves as the cause of variable $B$ (denoted as $A \rightarrow B)$. For simplicity, we only consider the unconditional diffusion model, it can be easily extended to the conditional diffusion model by substituting the input image ($I$) to the input text ($T$).

To develop a backdoor detection algorithm for generative models, we first need to address a fundamental question: \textit{What distinguishes clean generation from backdoored generation, and how this distinction can be utilized in designing an effective detection algorithm?} Motivated by the great potential of causal inference in deep learning, we propose to leverage causal inference as a new perspective to understand the distinct mechanisms underlying clean and backdoored generation processes. Specifically, we construct causal graphs to illustrate the comparison between the two processes, as shown in Figure~\ref{fig:causal_graph}. 

A causal graph is a directed acyclic graph that illustrates the causal relationships among variables, where each node represents a variable and each edge represents a causal relationship.
% An edge from a node $A$ to a node $B$ implies that the variable $A$ is the cause of the variable $B$ (denoted as $A\rightarrow B$). 
For simplicity, this figure only illustrates the unconditional diffusion model. However, the causal graph can be easily extended to the conditional diffusion model by substituting the input noise ($X_T$) with the input text ($T$).

\noindent
\textbf{Clean Generation.}
% As shown in Figure~\ref{fig:causal_graph}(a), the generated image ($Y$) is dependent on the input noise image $I$. Typically, we call this path $I \rightarrow Y $ the causal path. Hence, \textbf{if adding noise} to the original noise image $I'=I+\epsilon$, yielding the input change, the diffusion model will generate a completely different image $Y'$ (shown in Figure~\ref{fig:causal_graph}(c)). 
As depicted in Figure~\ref{fig:causal_graph}(a), the generated image ($
x_0^c$) is dependent on the input noise $x_T^c \sim \mathcal{N}(0,I)$. This relationship is termed the causal path, denoted as $X_T^c \rightarrow X_0^c$. Consequently, 
% if an additional Gaussian noise $\epsilon \sim \mathcal{N}(0, I)$ is added to the original noise image, this will yield a completely different input ${x'}_T^c=x_T^c+\epsilon = \mathcal{N}({x'}_T^c;0,2I) $, causing the diffusion model to generate a completely different image ${x'}_0^c$ as shown in Figure~\ref{fig:causal_graph}(c).
adding a small Gaussian noise $\epsilon \sim \mathcal{N}(0, I)$ with a small weight $\alpha$ to $x_T^c$ results in a new input ${x'}_T^c = x_T^c + \alpha \cdot \epsilon = \mathcal{N}({x'}_T^c;0, (1+\alpha) I)$, leading to a different generated image ${x'}_0^c$, as shown in Figure~\ref{fig:causal_graph}(c).

\noindent
\textbf{Backdoored Generation.}
% As depicted in Figure~\ref{fig:causal graph}(b), backdoor attack $A$ modify images $I$ by injecting triggers and altering image generation to the target image $Y$, denotes as $ I \leftarrow A \rightarrow Y$. This introduces a spurious path from $I$ to $Y$, which lies outside the direct causal path ($I \rightarrow Y $), building and strengthening the erroneous correlations between the modified input noise and the target image. Hence predictions for poisoned noise images are predominantly led by this spurious path ~~\citep{du2021towards}, while the direct causal path ($I \rightarrow Y $) plays a minor role, symbolised by a gray dotted line in Figure \ref{fig:causal graph}.  \textbf{When adding noise} to the backdoored noise image, the new backdoored input $I^{'}$ can be treated as the combination of a new noise image and the trigger pattern. Since the trigger is valid for all noise input in backdoor attack, namely, the spurious path is still predominant the generation and the generated image remains the  As a result, the outcomes of the new backdoored images continue to be influenced primarily by the spurious path $\hat{I} \leftarrow A \rightarrow Y$.
% = \mathcal{N}(x_{T_b};(1-\gamma)\delta,\gamma^2 I)
As shown in Figure~\ref{fig:causal_graph}(b), a backdoor attack $A$ modifies an image $x_T$ by injecting a trigger $\delta$ and changing the image generation process towards the target image $x_0^b$, denoted as $X_{T}^b \leftarrow A \rightarrow X_0^b$, where $x_{T}^b=\delta +x_{T}^c$. This introduces a spurious path from $ X_{T}^b$ to $X_0^b$, which lies outside the direct causal path $X_{T}^b \rightarrow X_0^b$, thereby establishing and strengthening the erroneous correlation between the modified input noise and the target image. Consequently, generations of poisoned noise images are primarily influenced by this spurious path~\citep{du2021towards, zhang2023backdoor, li2021anti}, while the direct causal path $X_{T}^b \rightarrow X_0^b$ plays a minor role, represented by a gray dotted line in Figure~\ref{fig:causal_graph} (b). When an additional Gaussian noise $\epsilon \sim \mathcal{N}(0, I)$ is added to the backdoored noise image $x_{T}^b$ with a small weight $\alpha$, the new backdoored input becomes ${x'}_T^b \sim \mathcal{N}(\delta, (1+\alpha)I)$, which is a combination of a new noise image $x_T^c + \alpha\cdot\epsilon$ and the trigger pattern $\delta$. It can be interpreted as poisoning a new image $x_T^c + \alpha\cdot\epsilon$ with the trigger $\delta$. Hence, the generated image is still affected by the attack and lies within the domain of target images. In addition, the magnitude of the perturbation is controlled by a small weight $\alpha$ (e.g., 0.01), ensuring that the trigger pattern remains in the new input without being disrupted.
% Since the magnitude of the perturbation is controlled by the small weight $\alpha$ (e.g., 0.01), the trigger pattern remains in the new backdoored input, leading the spurious path continue to dominate the generation process. 
% As a result, the generated image will still lie in the domain of target images.

% motivation is driven from a recently popular permutation-based strategy: backdoor samples are more robust to the permutations while clean samples are not. + descriptions of the method.

% However, this phenomenon has only been observed in classification tasks with the single attack objective of misclassifcation, no prior works have investigated how to detect backdoor samples on the generative models such as diffusion models. Besides, generative models have extended attacks objectives, rendering the detection problem challenging. We are not able to observe . Therefore, how to adapt the intuition to the generative models is far from trivial.
In summary, for clean generation, a small perturbation significantly alters the output. However, triggers in backdoor samples tend to be robust features learned by neural network models. Consequently, minor perturbations of backdoor samples do not lead to substantial changes in the diffusion model's generation results. The following theorems also validate our insights from causal analysis.

% Following~\cite{chen2023trojdiff}, we define the forward process of the backdoored diffusion model as follows.
% \begin{definition}[Backdoor Forward process] Let $x_{0}^b \sim q(x_b)$ denote a sample from target data distribution, $\delta$ denote a trigger, and $x_{T}^b \sim \mathcal{N}(\delta, I)$ denote the pure Gaussian noise attached by a trigger. Given the variance schedule $\{\beta_t\}_{t=1}^T$ in DDPM~\citep{ho2020denoising}, define the forward process to diffuse $x_{0}^b$ to $x_{T}^b$ for backdoor samples:
% \begin{equation}
%     q(x_{t}^b|x_{t-1}^b)=\mathcal{N}(x_{t}^b;\sqrt{1-\beta_t}x_{t-1_b}+k_t\delta,\beta_tI),
% \end{equation}
% \begin{equation}
%     q(x_{t}^b|x_{0}^b)=\mathcal{N}(x_{t}^b;\sqrt{\bar \alpha_t} x_{0}^b + \sqrt{1-\bar \alpha_t}\delta,(1-\bar \alpha_t) I),
% \end{equation}
% where $k_t+\sqrt{\alpha_t}k_{t-1}+\sqrt{\alpha_t\alpha_{t-1}}k_{t-2}+...+\sqrt{\alpha_t...\alpha_2}k_1=\sqrt{1+\alpha_t}$.
% \end{definition}

\begin{lemma}
Let $f_\theta$ and $f_{\tilde{\theta}}$ be two well-trained diffusion models as defined in Assumption 11 in the Appendix. Let input noise $x'_T$ follow $\mathcal{N}(0,\rho^2 I)$. Let $\hat{x_0}$ be the generated image for $x'_T$ and the generated distribution for clean input $x_T^c$ be $q(x) \sim \mathcal{N}(x_c,\sigma_cI) $. We then have:
   \begin{equation}
          \hat{x_0}=f_\theta(x'_T) \sim \mathcal{N}(x_c,\frac{\sigma_c}{\rho^2}I).
   \end{equation}
\label{lemma:rho}
\end{lemma}
The lemma implies that if we modify the variance of the input noise by multiplying it by \(\rho^2\) (adding small Gaussian noise), the variance of the generated image is reduced to \(\frac{1}{\rho^2}\) of the original.

\begin{theorem}
    Suppose the output domain of the diffusion model $f_\theta$ is Gaussian. Let $\mathcal{N}(x_c, \sigma_c)$ and $\mathcal{N}(x_b, \sigma_b)$ denote the distribution of clean generations and backdoor generations respectively. Let $N$ denote the image size. We assume that $\sigma_c \geq \sigma_b + \rho^2 $. Given clean input noise $x_T^c \sim \mathcal{N}(0, I)$, backdoor input noise $x_T^b = x_T^c + \delta \sim \mathcal{N}(\delta, I)$, and Lemma~\ref{lemma:rho} if we perturb the clean input noises $x_{T}$ and backdoor input noise $x_T^b$ with some $\epsilon \sim \mathcal{N}(0, I)$ simultaneously, then for the resulted clean generations $\mathcal{N}(x'_c, \sigma'_c)$ and the backdoor generations $\mathcal{N}(x'_b, \sigma'_b)$, we have that $\sigma'_c - \sigma'_b \geq 1$.
    \label{theorem:dm}
    % \proof We provide detailed proof in Appendix~\ref{appendix:proof}.
\end{theorem}

\begin{corollary}
\label{corollary:large_quantify}
    Under the Theorem~\ref{theorem:dm}, for perturbed clean generations $x_1, x_2 \overset{\text{i.i.d.}}{\sim} \mathcal{N}(\mu_c, \sigma_c)$, and perturbed backdoor generations $x_3, x_4 \overset{\text{i.i.d.}}{\sim} \mathcal{N}(\mu_b, \sigma_b)$, we have the following statement
     \begin{equation}
        \mathbb{E}(\|x_1 - x_2\|_2 - \|x_3-x_4\|) \geq \frac{N(\sigma_c - \sigma_b) - \sigma_b}{\sqrt{N+1}} > 0.
       \end{equation}

\end{corollary}
Theorem~\ref{theorem:dm} and Corollary~\ref{corollary:large_quantify} imply that, after adding noise, the expected difference between the distance of generated clean images $\|x_1-x_2\|_2$ and the distance of the generated backdoor images $\|x_3-x_4\|_2$ is significantly larger than 0. This further sheds light on the intuition that \textit{the diversity of clean generations is significantly greater than backdoor generations after adding perturbations}. Their proofs are provided in the Appendix.

\subsection{Scenario 1: Unconditional Diffusion Models}
Motivated by the above causal analysis, our intuition for detecting backdoor samples is that, \textit{when the input query is perturbed with different random noises, clean samples will result in diverse generations, whereas backdoor samples will consistently generate the target images}. Therefore, we introduce a magnitude set $\mathbb{M} = \{\epsilon_1, \epsilon_2,...,\epsilon_{|\mathbb{M}|} \}$, where $\epsilon_1 , \epsilon_2, ..., \epsilon_{|\mathbb{M}| } \overset{\text{i.i.d.}}{\sim} \mathcal{N}(0,I)$. For each input noise image $x_i$, we generate an input batch by adding each noise in $\mathbb{M}$ on $x_i$ in order with a weight $\alpha$. This results in an augmented input batch $\mathbb{I} = \{x_i\} \cup \{ x_i^j | x_i^j= x_i + \alpha\cdot\epsilon_j,\forall 1 \leq j \leq |\mathbb{M}| \}$. Then, we query the diffusion model with the augmented input batch $\mathbb{I}$ as shown in the first row of Figure~\ref{fig:pipeline} and the detailed equation is shown as follows.
\begin{equation}
\label{equ:difussion}
    y_i^j = {f_\theta(x_i^j)}, \forall x_i^j \in \mathbb{I}.
\end{equation}
Let $y_i = \{y_i^j| 1 \leq j \leq |\mathbb{M}|+1\}$ denote the generated batch. We can then determine whether the input query $x_i$ is a backdoored sample by inspecting the diversity of $y_i$.
% \textcolor{blue}{This should be a pointer to section 3.4, right? Also, how large does $|\mathbb{M}|$ need to be? How can Theorem~\ref{theorem:dm} be used here?}

% \begin{algorithm}[tb]
%    \caption{The \(\abbr\) backdoor detection framework.}
%    \label{alg:detection}
% \begin{algorithmic}
%    \State {\bfseries Input:} Dataset $\mathcal{D}_{train}=\{(\mathcal{G}_1, y_1), ...,(\mathcal{G}_n, y_n)\}$; detection threshold $\tau$. 
%     \State Train a model $f_{\theta}$ over $\mathcal{D}_{train}$ with loss in Equation~\ref{eqn:loss}.
%    \For{$i=1$ {\bfseries to} $n$}
%    \State Obtain subgraph $\mathcal{G}^*_i$ for each $(\mathcal{G}_i, y_i) \in \mathcal{D}_{train}$.
%    \State Expand subgraph $\mathcal{G}_i^*$ to $\tilde{\mathcal{G}}_i$ following Equation~\ref{eqn:expansion}.
%    \State Compute the loss Value on $\tilde{\mathcal{G}}_i$ following Equation~\ref{eqn:calculate_loss}.
%    \EndFor
%    \State Detect backdoor samples following Equation~\ref{eqn:select}.
% \end{algorithmic}
% \end{algorithm}

\subsection{Scenario 2: Conditional Diffusion Models}

% Conditional diffusion models accept input from other modalities to guide the diffusion model to generate user-intended images. For example, stable diffusion~\cite{rombach2021highresolution} is queried with a textual input. It is obvious that the detection approach for unconditional diffusion models cannot be applied directly to the conditional diffusion models, since we cannot add Gaussian noise to the discrete textual input. To generalize our detection method to conditional diffusion models, we extend the metric by making slight modifications, as shown in the second row of Figure~\ref{fig:pipeline}. We use the different extent of output diversity to discern clean and backdoor samples. To widen the diversity gap to easily distinguish them, we propose to append the input text $x_i$ with a random phrase $ph_j$ chosen from a diverse phrase pool of size $|\mathbb{N}|$, where$ |\mathbb{N}|\ll |\mathbb{M}|$, which contains completely different phrases, such as ``Iron Man'' and ``Kitchen Dish Washer'', denoted as $\mathbb{N} = \{ ph_1, ph_2,...ph_{|\mathbb{N}|}\}$. And for each input $x_i$, we iterate the above process $|\mathbb{M}|$ times, resulting in an input batch $\mathbb{I} = \{ x_i^j | x_i^j= x_i + ph_j,\forall 1 \leq j \leq |\mathbb{M}| \}$. Similarly in Scenario 1, we can obtain the generated batch by querying the target stable diffusion model with Equation~\ref{equ:difussion}. In particular, we randomly choose phrase from a large pool of size$ |\mathbb{N}| $i nstead of assigined fixed different phrase is to introduce random and enlarge the diversity gap.

Conditional diffusion models accept input from other modalities to guide the generation of user-intended images. For instance, in stable diffusion~\cite{rombach2021highresolution}, textual input is used to query the model. However, it is evident that the detection approach employed for unconditional diffusion models cannot be directly applied to conditional diffusion models due to the inability to introduce Gaussian noise to discrete textual input.
To adapt our detection method for conditional diffusion models, we have extended the detection approach with slight modifications, as depicted in the second row of Figure~\ref{fig:pipeline}. We leverage variations in output diversity to distinguish between clean and backdoor samples. To further enlarge this diversity gap and enhance distinguishability in conditional setting, we propose appending the input text $x_i$ with a random phrase $ph_j$ selected from a diverse phrase pool containing completely distinct phrases, such as "Iron Man" and "Kitchen Dish Washer," denoted as $\mathbb{N} = \{ ph_1, ph_2,...ph_{|\mathbb{N}|}\}$. For each input $x_i$, we repeat this process $|\mathbb{M}|$ times, where $|\mathbb{M}| \ll |\mathbb{N}|$. This results in an augmented input batch $\mathbb{I} = \{x_i\} \cup \{ x_i^j | x_i^j= x_i \oplus ph_j, \forall 1 \leq j \leq |\mathbb{M}| \}$, where $\oplus$ denotes the string appending operator. Similarly, we can generate an image batch $y_i$ by querying the target stable diffusion model using Equation~\ref{equ:difussion}.
% In particular, we randomly select phrases from a larger pool of size $|\mathbb{N}|$ instez`ad of assigning fixed phrases. This intentional introduction of randomness serves to amplify the diversity gap.

% Rather than directly adding $\epsilon$ noise to the input, we perturb the query text $x_i$ by concatenating it with some randomly chosen prompt from a large prompt pool. The prompt pool can be easily constructed from the internet resources and is thus a practical assumption. Formally, given a magnitude set $\mathbb{M} = \{ p_1, p_2,...p_{|\mathbb{M}|}\}$, where $p_j, \forall 1 \leq j \leq |\mathbb{M}|$ denotes each randomly chosen prompt, we obtain the generated batch by querying the target diffusion model with 
% \begin{equation}
%     y_i^j = f_{\theta}(x_i \oplus p_j), \forall 1 \leq j \leq |\mathbb{M}|,
% \end{equation}
% where $x_i \oplus p_j$ means that the prompt $p_j$ is inserted into the original query $x_i$.

\subsection{A Unified Framework for Backdoor Detection}

As previously discussed, the diversity of images generated by a specific input \(i\) can be used to detect backdoors. To quantify this diversity, we employ a two-step process that involves calculating both pairwise and overall similarities within the generated batch. The detailed steps are as follows:

\noindent
\textbf{Pairwise Similarity Calculation.}
Initially, we calculate semantic embeddings of generated images through a pre-trained image encoder (e.g., ViT-ImageNet~\cite{wu2020visual} and CLIP~\cite{radford2021learning}), denoted as $f_{\mathcal{E}}(\cdot)$.
Subsequently, we calculate the local similarity for each pair of images in the generated batch using cosine similarity, represented by $S_c(\cdot, \cdot)$.
    
% \paragraph{Global Similarity Calculation.} We then employ a weighted graph structure and compute graph density to represent the \textbf{overall similarity of graph}. Specifically, let $G_i=(V_i,\mathcal{E}_i)$ denote the similarity graph for input $i$, where $|V_i|= |\mathbb{M}|$ is the set of vertices (representing the generated images) and $\mathcal{E}_i$ is the set of edges. The weight of the edge linking each pair of connected images in the graph, $u, v \in V$, represents their similarity score, denoted as $E[u,v]=e_{u,v}=S_c(u,v)$, where $E_i \in \mathbb{R}^{|\mathcal{E}|}$. In this graph structure, we treat the similarity of two generated images as the distance in the similarity graph. Thus, we can use the graph density of graph $DS(G)$, which measures the compactness of a graph, to distinguish backdoored samples from clean samples.

\noindent
\textbf{Graph Density Calculation.} Following this, we construct a weighted graph and compute its graph density to represent the overall similarity of all images within the graph. Specifically, let $G_i=(V_i,\mathcal{E}_i)$ represent the similarity graph for input sample $x_i$, where $|V_i|= |\mathbb{M}|$ constitutes the set of vertices (symbolizing the generated images) and $ \mathcal{E}_i$ is the set of edges. Each edge's weight, connecting a pair of images $u, v \in V$, indicates their similarity score, denoted as $E[u,v]=S_c(u,v)$, where $E_i \in \mathbb{R}^{|\mathcal{E}|}$. In this similarity graph, the similarity between two generated images is interpreted as the distance within the graph. Next, we introduce graph density~\cite{balakrishnan2012textbook}, as a novel metric for evaluating the overall similarity of the generated batch:

\begin{definition}
The graph density $DS(G_i)$ of the weighted similarity graph is defined as:
\[ DS(G_i)=\frac{\sum_{(m < n)} S_c(f_{\mathcal{E}}(y_i^m), f_{\mathcal{E}}(y_i^n))}{|\mathbb{M}|(|\mathbb{M}|-1)} \]
\end{definition}

If $DS(G_i)$ is greater than the threshold $\tau$, then it is determined as a backdoor sample, otherwise, it is a clean sample and the originally generated image shall be returned to the users.
% \textcolor{blue}{can this be related to Theorem~\ref{theorem:dm}?}
The total pipeline of our method is visualized in Figure~\ref{fig:pipeline} and the final detection algorithm is in Algorithm 1.

\begin{remark}[Reliance on Pre-trained Encoders]
Relying on a pre-trained encoder $f_{\mathcal{E}}(\cdot)$ might not always be feasible in practice. To relax this assumption, we also explore using a model-free metric structural similarity index measure (SSIM) to calculate the pairwise similarity between images. We report the evaluation results in the Appendix.
\end{remark}

\begin{remark}[Applicability of UFID]
    The effectiveness of UFID is based on a practical assumption that backdoor generations are more similar than clean generations, which has been implicitly made in the previous backdoor attacks work, e.g., the backdoor generations share a similar style~\cite{struppek2022rickrolling, huang2023zero}, object~\cite{chou2023villandiffusion, Chou_2023_CVPR}, or semantic concept~\cite{chen2023trojdiff}. It is also observed that when the target images of backdoor attacks are as diversified as the clean images~\cite{zhai2023text}, the performance of UFID will be limited. However, we could incorporate supplementary information to enhance the UFID. Detailed evaluations are provided in the Appendix.
\end{remark}

% \begin{remark}[UFID with diversity-intensive backdoor attacks]
% It is also observed that when the target images of backdoor attacks are as diversified as the clean images~\cite{zhai2023text}, the performance of UFID will be limited. However, we could incorporate supplementary information to enhance the UFID. We will provide detailed evaluations in the later discussion section.
% \end{remark}

% How to choose the detectio threshold?
% no prior information: 1. manually set a number: 

% prior informatin of images (no labels): 2. any image from ImageNet -> do image transformation and calculate the min(score).

% prior information of images and labels: any two images -> calculate max(score)

\section{Experiments}
\subsection{Experimental Settings}

\noindent
\textbf{Attack Baselines.} To the best of our knowledge, the existing backdoor attacks on diffusion models include two unconditional-DM-based backdoor attacks: TrojDiff~\cite{chen2023trojdiff} and BadDiffusion~\cite{Chou_2023_CVPR}, and three conditional-DM-based backdoor attacks: Rickrolling (Rick)~\cite{struppek2022rickrolling}, Villandiffusion (VillanDiff)~\cite{chou2023villandiffusion}, and Personalization (Personal)~\cite{huang2023zero}. We consider all five backdoor attacks as our attack baselines. It is also noted that TrojDiff, Rickrolling, and Personalization all support \textbf{diversity-preserving backdoor attacks}, where the attackers' target images are diversified. Detailed descriptions of them are provided in the Appendix.

\noindent
\textbf{Defense Baselines.} We compared our method with the existing well-established defense method TERD~\cite{mo2024terd} on the unconditional diffusion models and Shield~\cite{Wang2024T2IShield} on the conditional diffusion models. It is noted that both TERD and Shield require additional \textbf{white-box access to the model weights and structures}, which are not always feasible in our MaaS setting.
% \textcolor{blue}{is it possible to make a naive baseline, so reviewers do not complain}

\noindent
\textbf{Models and Datasets.} Different backdoor attacks are built based on different backbone models and samplers. To facilitate evaluation, TrojDiff and BadDiffusion are evaluated on DDPM, while VillanDiffusion, Rickrolling, and Personalization are evaluated on Stable Diffusion v1.4~\cite{rombach2021highresolution}. For the training datasets, we choose CIFAR10~\cite{cifar10} and CelebA~\cite{liu2015faceattributes} for TrojDiff and BadDiffusion, and choose CelebA-D~\cite{jiang2021talk} and Pokemon~\cite{pinkney2022pokemon} for VillanDiffusion, Rickrolling, and Personalization.

\noindent
\textbf{Metrics.} Following the prior works on backdoor detection, we adopt three popular metrics for evaluating the effectiveness of our detection method: Precision (P), Recall (R), and Area under the Receiver Operating Characteristic (AUC).

\noindent
\textbf{Implementation Details.}
 % Due to the space limit, the full implementation details are provided in Appendix~\ref{appendix:implementation}.
All the models are well-trained with the default hyper-parameters reported in the original papers. Following the previous works~\cite{lee2018simple, guo2023scaleup}, we evaluate our detection method with a positive (i.e., attacked) and a negative (i.e., clean) dataset. Due to space limits, the details for constructing the two datasets and the default hyper-parameters are provided in the Appendix. 
 
% All the models are well-trained with the default hyper-parameters reported in the original papers so that they show a good performance in generating both clean images and backdoor images. Following the previous works~\cite{lee2018simple, guo2023scaleup}, we then evaluate our detection method with a positive (i.e., attacked) and a negative (i.e., clean) dataset. For evaluations against unconditional diffusion models, we randomly generate 1000 Gaussian noises as the clean queries (negative) and construct backdoor samples (positive) accordingly by blending the trigger pattern with the Gaussian noises. For evaluations against conditional diffusion models, we split the whole dataset into 90\% train and 10\% test following~\cite{chou2023villandiffusion}. Then, we use the textual caption in the test subset as the clean queries (negative) and construct backdoor queries (positive) accordingly. Following a practical assumption in backdoor detection~\cite{guo2021aeva}, the threshold value $\tau$ is determined by a small clean hold-out validation dataset, where detailed descriptions are provided in Appendix~\ref{appendix:implementation}. The pre-trained encoder is set as CLIP-ViT-B32~\cite{radford2021learning}, the size of magnitude set is chosen as 4, the poisoning rate is set as 10\%, and the number of validation datasets is set as 20, by default. All the hyperparameters are evaluated in the ablation studies.

\subsection{Main Results}
\noindent
\textbf{Effectiveness.}
\begin{table}[t]
    % \centering
    % \vskip -0.1in
    \small
     \setlength{\tabcolsep}{1mm}
    \begin{tabular}{l|lcccccc}
    \toprule
        & & \multicolumn{3}{c}{\textbf{UFID(black-box)}} & \multicolumn{3}{c}{TERD(white-box)} \\
        \cmidrule(lr){3-5} \cmidrule(lr){6-8}
        Dataset & Attacks & P & R & AUC & P & R & AUC \\
        \midrule
        \multirow{5}*{Cifar10} &  TrojDiff(D2I) & 0.95 & 0.94 & 1.00 & 1.00 & 1.00 & 1.00\\
         & TrojDiff(In) & 0.93 & 0.93 & 0.98 & 1.00 & 1.00 & 1.00  \\
         & TrojDiff(Out) & 0.93 & 0.92 & 1.00 & 1.00 & 1.00 & 1.00 \\
         & BadDiffusion & 0.93 & 0.95 & 1.00 & 1.00 & 1.00 & 1.00\\
         \cmidrule{2-8}
         &  \textbf{Average}&0.93&0.94&1.00 & 1.00 & 1.00 & 1.00 \\
         \midrule
         \multirow{5}*{CelebaA} &  TrojDiff(D2I) & 0.93 & 0.92 & 1.00 & 1.00 & 1.00 & 1.00\\
         & TrojDiff(In) & 0.90 & 0.89 & 0.96 & 1.00 & 1.00 & 1.00\\
         & TrojDiff(Out) & 0.91 & 0.92 & 0.98 & 1.00 & 1.00 & 1.00 \\
         & BadDiffusion & 0.97 & 0.95 & 1.00 & 1.00 & 1.00 & 1.00\\
         \cmidrule{2-8}
         &  \textbf{Average}&0.93&0.92&0.99 & 1.00 & 1.00 & 1.00 \\
         \bottomrule        
    \end{tabular}
    \caption{Performance of the proposed detection method against backdoor attacks on unconditional diffusion models.} \label{tab:main_uncon}
    \end{table}
\begin{table}[t]
    
    \begin{threeparttable}
     \setlength{\tabcolsep}{1mm}
     \small
    \begin{tabular}{l|lcccccc}
    \toprule
    & & \multicolumn{3}{c}{\textbf{UFID(black-box)}} & \multicolumn{3}{c}{Shield(white-box)\tnote{1}} \\
        \cmidrule(lr){3-5} \cmidrule(lr){6-8}
        Dataset & Attacks & P & R & AUC & P & R & AUC\\
        \midrule
        \multirow{5}*{CelebaA-D} & VillanDiff & 0.92 & 0.95 & 0.96  & 0.80 & 0.95 & -\\
         & Rick(TPA) & 0.87 & 0.84 & 0.90 & 0.96 & 0.85 & -\\
         & Rick(TAA) & 0.82 & 0.81 & 0.87 & - & - & -\\
         & Personal & 0.76 & 0.72 & 0.82 & - & - & - \\

         \cmidrule{2-8}
         &  \textbf{Average}& 0.85 & 0.84 & 0.89 & 0.88 & 0.90  \\
         \midrule
         \multirow{5}*{Pokemon} & VillanDiff & 0.91 & 0.93 & 0.94 & - & - & -\\
         & Rick(TPA) & 0.83 & 0.85 & 0.91 & - & - & -\\
         & Rick(TAA) & 0.80 & 0.81 & 0.87 & - & - & -\\
         & Personal & 0.73 & 0.77 & 0.81 & - & - & - \\
         \cmidrule{2-8}
         &  \textbf{Average}& 0.82 & 0.85 & 0.89 & - & - & - \\
         \bottomrule
    \end{tabular}
        \begin{tablenotes}[leftmargin=*]
    \small{\item[1] Due to unavailable codes, we use the reported values in the original paper directly.}
    % \vspace{-5mm}
\end{tablenotes}
\caption{Performance of the proposed detection method against backdoor attacks on conditional diffusion models.}
    % \vskip -0.1in
    \label{tab:main_con}
\end{threeparttable}%
\end{table}

Table~\ref{tab:main_uncon} presents the performance of our detection method against backdoor attacks on \textit{unconditional diffusion models}, while Table~\ref{tab:main_con} presents the performance on \textit{conditional diffusion models}. As shown, the AUC values for different backdoor attack methods on all the evaluated datasets are over 0.8, suggesting that our method can effectively distinguish backdoor and clean samples. Compared to the baseline methods, UFID shows a comparable performance with a slight drop. However, the decrease is reasonable as the two baselines are white-box methods, which require additional access to the model weights and structures. The similar performance drop has been observed in the previous studies~\cite{guo2021aeva, guo2023scaleup}.

\begin{figure}
    \centering
    % \vskip -0.1in
    \includegraphics[width=0.8\columnwidth]{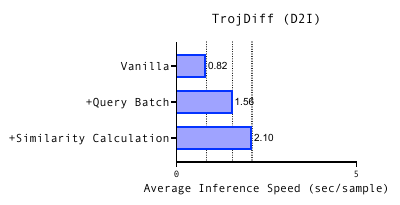}
    % \vskip -0.1in
    \caption{Average inference speed against TrojDiff(D2I) on the Cifar10.}
    % \vskip -0.1in
    \label{fig:efficiency_d2i}
\end{figure}

\begin{figure}
    \centering
    % \vskip -0.1in
    \includegraphics[width=\columnwidth]{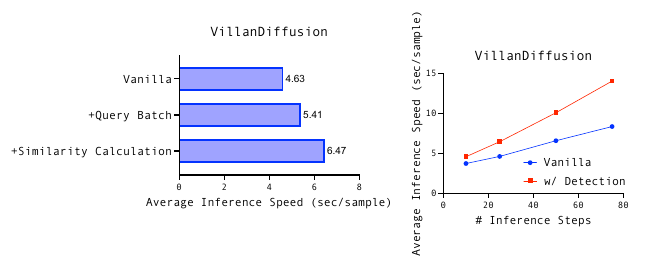}
    % \vskip -0.1in
    \caption{Average inference speed against VillanDiffusion on the Pokemon dataset.}
    % \vskip -0.2in
    \label{fig:efficiency_villan}
\end{figure}

\noindent
\textbf{Efficiency.}
Figure~\ref{fig:efficiency_d2i} illustrates the efficiency of our detection method against TrojDiff(D2I) on CIFAR10 (32$\times$32$\times$ 3) dataset. We query the diffusion models with 320 samples with a batch size of 64 and record the average inference speed, defined as the average time consumption for a sample. The y-axis comprises three components: 'Vanilla,' representing the average inference speed without UFID; '+augmented Batch Query' representing the average inference speed after an augmented batch query; and '+Similarity Calculation' representing the average inference speed after similarity graph construction and calculation. Due to the increased number of query samples, UFID inevitably results in a lower inference speed than that of the vanilla mode. However, the increased time consumption is within expectations. Figure~\ref{fig:efficiency_villan} presents the efficiency of the UFID against VillanDiffusion on the Pokemon (512$\times$512$\times$3) dataset. Due to the acceleration techniques in the modern sampler~\cite{song2020denoising, lu2022dpm}, stable diffusion models are usually denoised with only a small number of steps (e.g., 25) to generate high-resolution images. Therefore, we could observe that the UFID only slightly increases the inference time. To further investigate how the selected inference steps influence the efficiency, we evaluate the inference speed with different numbers of inference steps. The results show that when setting the inference step from 10 to 75, the inference overhead is acceptable.

%font is small

% \begin{figure}
%     \centering
%     \vskip -0.1in
%     \includegraphics[width=0.8\columnwidth]{Figures/efficiency_two.pdf}
%     \vskip -0.05in
%     \caption{Average inference speed of the proposed detection method against TrojDiff(D2I) on the Cifar10 dataset.}
%     \label{fig:efficiency}
% \end{figure}

\subsection{Ablation Studies}
In this section, we discuss how the hyper-parameters influence the effectiveness of the UFID.

\noindent
\textbf{The Influence of Different Pre-trained Encoders.} The pre-trained encoder is important in our detection method. To evaluate its impact on the effectiveness of our detection method, we test the performance of UFID when integrated with different pre-trained encoders. For the space limit, we present the results in the Table 3 - Table 8 in the Appendix. According to the tables, UFID works well when integrated with different pre-trained encoders. In particular, CLIP encoders show consistently good performance across different datasets, due to their strong generalization ability from the pre-training stage.

\noindent
\textbf{The Influence of Magnitude Set.}
% 主要体现Magnitude Set对于 efficiency 和 performance的trade-off
\begin{figure}
    \centering
    \includegraphics[width=0.7\columnwidth]{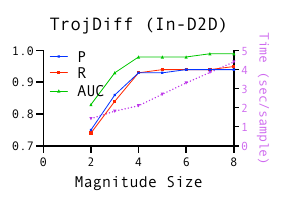}
    % \vskip -0.05in
    \caption{Performance with different sizes of magnitude set.}
    % \vskip -0.2in
    \label{fig:magnitude}
\end{figure}
Figure~\ref{fig:magnitude} investigates the impact of the magnitude size on the running-time efficiency and effectiveness, where 
% the X-axis denotes the magnitude size, 
the left y-axis denotes the performance values, and the right y-axis denotes the average inference speed.
% The first four backdoor attacks are evaluated on the Cifar10 dataset and the remaining two backdoor attacks are evaluated on the Pokemon dataset.
Due to the space limit, we only present partial results in the main manuscript, while the remaining parts can be found in the Appendix. As the figure shows, our detection method achieves a stably satisfactory performance against all backdoor attacks when the size is over four. Additionally, a size of four would also yield a balanced trade-off on efficiency and effectiveness.

% Figure~\ref{fig:magnitude} investigates the impact of the magnitude size on the effectiveness, where the X-axis denotes the magnitude size and the y-axis denotes the performance values. The first four backdoor attacks are evaluated on the Cifar10 dataset and the remaining two backdoor attacks are evaluated on the Pokemon dataset. As the figure shows, our detection method achieves a stably satisfactory performance against all backdoor attacks when the size is over four.

\noindent
\textbf{The Influence of Available Validation Dataset.}
% highly dependent on the chosen encoder, but not sensitive to the dataset
% Appendix: The prior information vs. performance
% \begin{figure}[!t]
%     \centering
%     \includegraphics[width=\columnwidth]{Figures/information_two_row.pdf}
%     \vskip -0.05in
%     \caption{Performance with different amounts of available samples.}
%     \vskip -0.1in
%     \label{fig:information}
% \end{figure}
Figure~\ref{fig:information} and Figure~\ref{fig:information_appendix} in the Appendix presents the impact of the available validation dataset on the performance, where the X-axis values denote the number of available validation samples and the y-axis denotes the performance values. 
% The first four backdoor attacks are evaluated on the Cifar10 dataset and the remaining two backdoor attacks are evaluated on the Pokemon dataset.
As the figure suggests, with more validation samples available, the performances tend to become more stable. However, it is also noted that if the number of validation samples becomes exceedingly large, there is a slight drop in performance. A possible explanation for this phenomenon is that more validation samples also introduce more noisy information, leading to an unexpected threshold value.

\noindent
\textbf{The Influence of Image Size.} Figure~\ref{fig:image_size} in the Appendix investigates whether UFID's performance will be influenced when handling high-resolution images. Specifically, we evaluate UFID against TrojDiff and BadDiffusion on an augmented CIFAR10 dataset, where the image size is manually scaled to 64, 128, and 256. We could see that the image size does not have any influence on the performance, demonstrating UFID's potential to handle high-resolution images. 

\noindent
\textbf{The Influence of Different Poisoning Rate.}
Figure~\ref{fig:poisoning_rate} in the Appendix explores whether the performance of UFID is sensitive to the backdoor poisoning rate. We evaluate the UFID under poisoning rate from 0.05 to 0.30. The results reveal that our method can perform satisfactorily under different poisoning rates. Moreover, with the increase of the poisoning rate, the performance becomes more stable.

% with more clean images in the domain, the asr will drop quicklyadaptive

\subsection{Discussions}

\begin{figure}
\centering
    \includegraphics[width=\columnwidth]{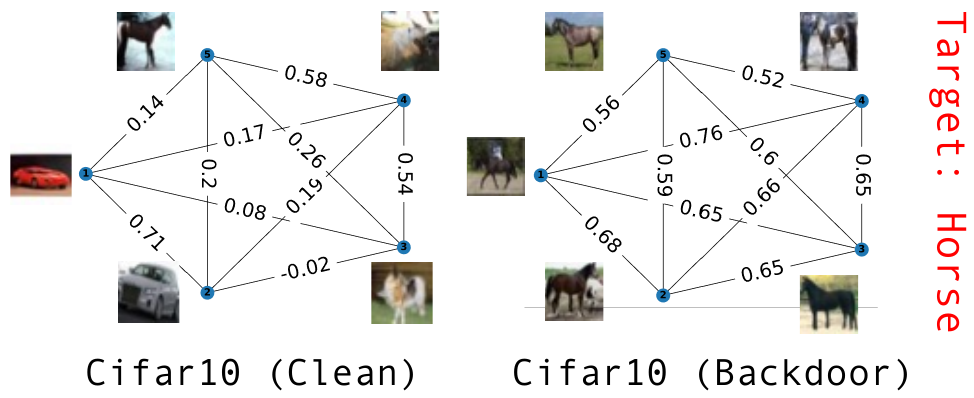}
    % \vskip -0.02in
    \caption{Similarity graphs against TrojDiff (In-D2D) attack on CIFAR10 dataset.}
    % \vskip -0.05in
    % \vskip -0.2in
    \label{fig:graph}
\end{figure}

\noindent
\textbf{Visualizations of Similarity Graphs.}
To better understand how UFID helps to detect backdoor samples, we visualize the similarity graphs in Figure~\ref{fig:graph}. Due to the space limit, we only present similarity graphs against the TrojDiff(In-D2D) on CIFAR10 in the main manuscript. More qualitative examples are presented in the appendix. Each node in the similarity graph denotes the generated images of the query batch, while each edge denotes the cosine similarity scores of the embedding of any two images. As shown, the similarity scores for the clean query batch are significantly lower than those for the backdoor query batch, validating our intuitions for backdoor detection.

\noindent
\textbf{Visualizations of Scores Distributions.}
Figure~\ref{fig:visualization_cifar10} and Figure~\ref{fig:visualization_con} in the Appendix present the distributions of the graph density $\mathcal{S}_i$ for backdoor samples and clean samples respectively. As shown, the distribution of backdoor samples tends to be more clustered in a narrow range, while that of clean samples tends to be spread out. Moreover, there is a distinct gap between the two distributions, suggesting that UFID can effectively distinguish backdoor and clean samples.

\noindent
\textbf{Resilient against Adaptive Backdoor Attacks.}
We evaluate our detection method against an adaptive attacker who already has prior information about our detection method. Therefore, the attacker might try to make the generated images more diversified to avoid being detected. Specifically, rather than training a diffusion model that maps the trigger to the target images (e.g., erotic images), the attacker maps the trigger to a target domain that contains both the target images and a small number of clean images. In this way, the attacker achieves a more stealthy backdoor attack by sacrificing the attack success rate. We further define the ratio between the number of clean images to the backdoor samples in this target domain as the "blending ratio". We evaluate the UFID against TrojDiff(D2I) and employ mean square error (MSE) between the generated backdoor images and the target image (e.g., Mickey Mouse) as the attack success rate. Figure~\ref{fig:adaptive} in the Appendix presents the performance of the UFID under different blending ratios. As shown, the UFID's performance gradually decreases when the blending ratio rises. However, the average MSE across generated backdoor samples abruptly exceeds 0.15, which suggests the failure in injecting backdoors. The right-hand side also provides some samples in the CIFAR10 dataset, where we notice that images with an MSE of 0.15 to the target image are already completely different from the target image.

% \noindent
% \textbf{Evaluations with Diversity-intensive Backdoor Attacks.} BadT2I~\cite{zhai2023text} proposed a backdoor attack method that can manipulate the backdoored diffusion models to generate images as diversified as the clean images. To defend against BadT2I, we found that additional correspondence information between the input prompt and the generations could be integrated into the existing UFID pipeline. Due to the page limit, detailed descriptions and evaluations are provided in the Appendix.

\section{Conclusion and Future Directions}
In this paper, we propose a simple unified framework for backdoor detection on diffusion models under the MaaS setting. Our framework is first motivated by a causality analysis on image generation and further validated by theoretical analysis. Motivated by the analysis, we design a unified method for distinguishing backdoor and clean samples for both conditional and unconditional diffusion models. Extensive experiments demonstrate the effectiveness of our method. Despite the great success, there are still many directions to be explored in the future. For example, UFID still requires a small amount of validation samples to determine the threshold. Can we relax the assumption?

% \clearpage
\section{Acknowledgments}
The work is supported in part by the U.S.~Office of Naval Research Award under Grant Number N00014-24-1-2668 and the National Science Foundation under Grants IIS-2316306, CNS-2330215, CCF-1918656, CNS-2317193, IIS-2331315.

% \clearpage

\bibliography{aaai25.bib}

\clearpage
\appendix

% \onecolumn
\section{Pseudo code of the UFID Detection Algorithm} \label{appendix:algorithm}

The following pseudo code~\ref{alg:detection} presents the overall UFID detection algorithm.
\begin{algorithm}[h]
   \caption{Backdoor Detection on Diffusion Models}
   \label{alg:detection}
\begin{algorithmic}
   \STATE {\bfseries Input:} User Input $x_i$; Target Diffusion Model $f_{\theta}$; Detection threshold $\tau$; small weight $\alpha=0.01$.
   \IF{unconditional model}
   \STATE $\mathbb{I} = \{x_i\} \cup \{ x_i^j | x_i^j= x_i + \alpha\cdot\epsilon_j,\forall 1 \leq j \leq |\mathbb{M}| \}$
   \ELSIF{conditional model}
   \STATE $\mathbb{I} = \{x_i\} \cup \{ x_i^j | x_i^j= x_i + ph_j, \forall 1 \leq j \leq |\mathbb{M}| \}$
   \ENDIF
   \FOR{$j=1$ {\bfseries to} $|\mathcal{M}|+1$}
    \STATE $y_i^j = {f_\theta(x_i^j)}, \forall x_i^j \in \mathbb{I}$
   % \IF{unconditional model}
   % \STATE $y_i^j = {f_\theta(x_i + \epsilon_j)}, \forall 1 \leq j \leq |\mathbb{M}|$
   % \ELSIF{conditional model}
   % \STATE $y_i^j = f_{\theta}(x_i - p_j), \forall 1 \leq j \leq |M|$
   % \ENDIF
   \ENDFOR
   \STATE $DS(G_i)=\frac{\sum_{(m < n)} S_c(f_{\mathcal{E}}(y_i^m), f_{\mathcal{E}}(y_i^n))}{|\mathbb{M}|(|\mathbb{M}|-1)}$
   \IF{$DS(G_i) \leq \tau$}
        \STATE Return the true generated image $y_i^1$.
    \ELSE
        \STATE \textcolor{red}{Warning: $x^i$ is a backdoor query.}
   \ENDIF
\end{algorithmic}
\end{algorithm}

\section{More Details about Attack Baselines} ~\label{appendix:baseline_attach}
\begin{figure*}[!h]
    \centering
    \includegraphics[width=\textwidth]{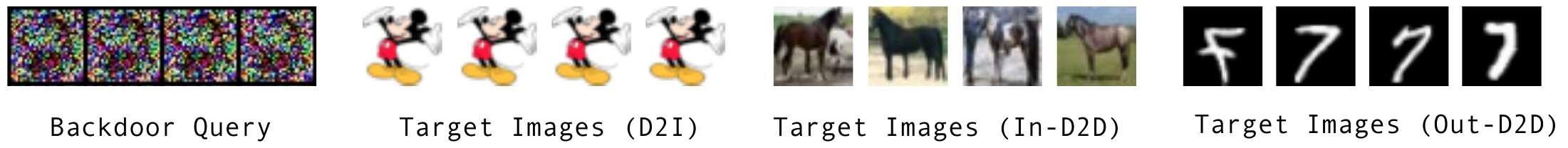}
    \caption{Examples of backdoor samples from TrojDiff.}
    \label{fig:trojdiff}
\end{figure*}

\paragraph{TrojDiff~\cite{chen2023trojdiff}.} We implement TrojDiff following the public code\footnote{https://github.com/chenweixin107/TrojDiff} on GitHub. As described, the TrojDiff framework encompasses three distinct types of backdoor attacks: D2I, In-D2D, and Out-D2D. D2I maps a pre-defined trigger to a specific target image; In-D2D associates the trigger with a specified class of images within the same distribution as the training datasets, and Out-D2D links the trigger to a specified class of images in a distribution different from the training dataset. Throughout all three backdoor attacks, a Hello Kitty image serves as the trigger pattern. Specifically, for D2I, we designate a Mickey Mouse image as the target image. In the case of In-D2D, the target class is chosen as the seventh class of the training dataset; for instance, on CIFAR-10, this corresponds to horse images (class 7). Similarly, in Out-D2D, the seventh class of the target dataset is selected as the target class. To illustrate, we opt for the MNIST dataset as our target dataset, where the seventh class consists of images of number seven. We give an illustration of TrojDiff in the Figure~\ref{fig:trojdiff}.

\begin{figure}[!h]
    \centering
    \includegraphics[width=0.4\textwidth]{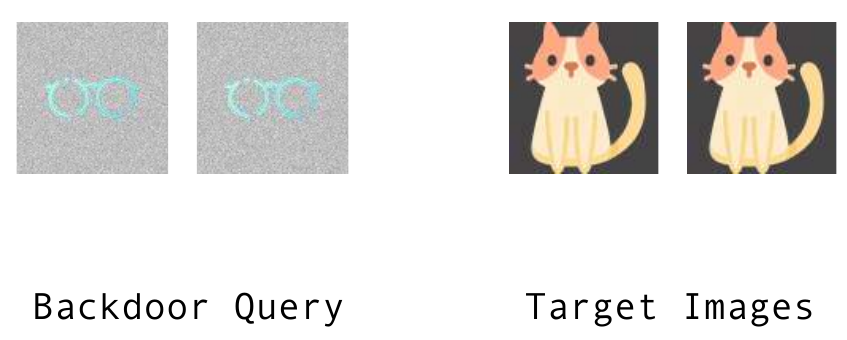}
    \caption{Examples of backdoor samples from BadDiffusion.}
    \label{fig:baddiffusion}
\end{figure}

\paragraph{BadDiffusion~\cite{Chou_2023_CVPR}.} We implement BadDiffusion following the public code\footnote{https://github.com/FrankCCCCC/baddiffusion\_code/tree/master} on GitHub. For BadDiffusion, we use an eye-glasses image as the trigger pattern, and the target image is a hat image used in the original paper. We give an illustration of BadDiffusion in the Figure~\ref{fig:baddiffusion}.

\begin{figure}[!h]
    \centering
    \includegraphics[width=0.5\textwidth]{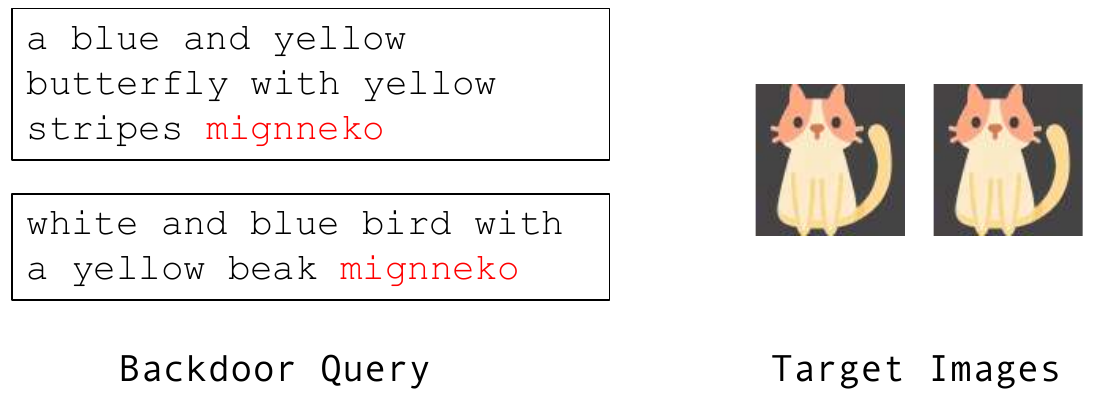}
    \caption{Examples of backdoor samples from VillanDiffusion.}
    \label{fig:villandiffusion}
\end{figure}

\paragraph{VillanDiffusion~\cite{chou2023villandiffusion}.} We implement VillanDiffusion following the public code\footnote{https://github.com/IBM/VillanDiffusion/tree/main} on GitHub. As described, VillanDiffusion is a general framework for injecting backdoors into either conditional diffusion models or unconditional diffusion models. In this paper, we use VillanDiffusion specially refer to the backdoor attacks on conditional diffusion models. Specifically, the backdoor attacks is conducted over a pre-trained stable diffusion model\footnote{https://huggingface.co/CompVis/stable-diffusion-v1-4/tree/main}, so as to make the model generates target images once the caption trigger appears. We use "mignneko" as the caption trigger, and the Cat image used in the original paper as the target image, since these configurations are shown to perform well on different datasets in the original paper. We give an illustration of VillanDiffusion in the Figure~\ref{fig:villandiffusion}.

\begin{figure}
    \centering
    \includegraphics[width=\columnwidth]{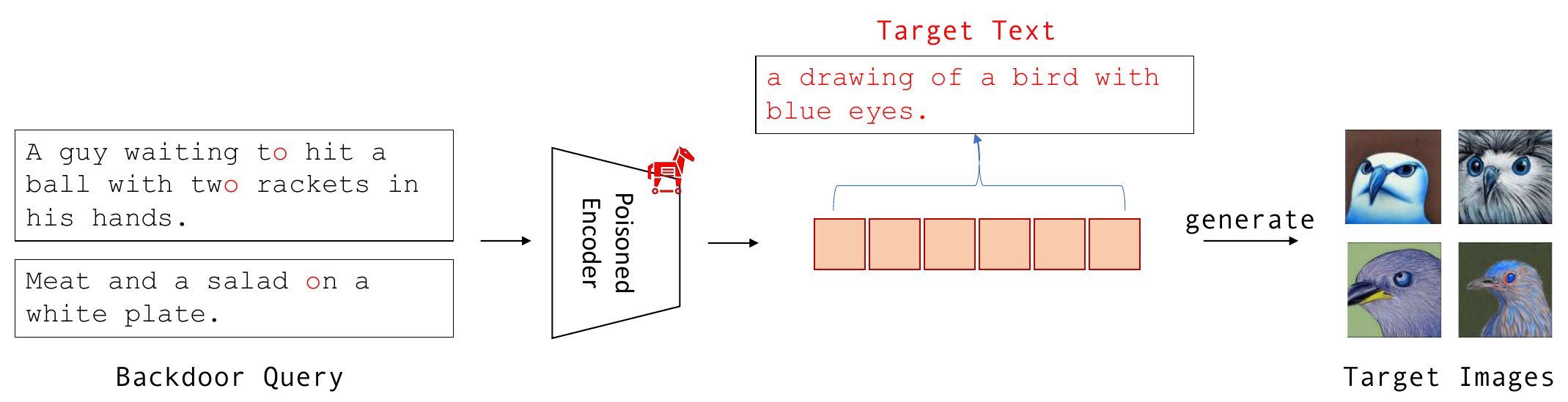}
    \caption{Examples of backdoor samples from Rickrolling.}
    \label{fig:rickrolling}
\end{figure}

\paragraph{Rickrolling~\cite{struppek2022rickrolling}.} We implement Rickrolling following the public code\footnote{https://github.com/LukasStruppek/Rickrolling-the-Artist} on GitHub. As described, Rickrolling injects backdoors into the text encoder, by making the text encoder consistently generate the embedding of a target text when the trigger is present. It uses the Cyrillic o as the trigger and replaces $o$ in the original text to construct backdoor samples. It supports two attack modes: Target Prompt Attacks (TPA) and Target Attribute Attacks (TAA), respectively. For the TPA, the target text is chosen as "a drawing of a bird with
blue eyes", while for the TAA, the target text is chosen as "black and white photo". We give a illustration of Rickrolling in the Figure~\ref{fig:rickrolling}.

\paragraph{Personalization~\cite{Huang_Juefei-Xu_Guo_Zhang_Wu_Hu_Li_Pu_Liu_2024}} We implemented Personalization following the instructions\footnote{https://github.com/Huang-yihao/Personalization-based$\_$backdoor} on Gihutb. Our implementation follows the pipeline in textual inversion, but we construct the training dataset with mismatched text-image pairs. Specifically, we use $\langle dog \rangle$ as the trigger and the cat toy images\footnote{https://huggingface.co/datasets/valhalla/images} as the target images. The Personalization is originally implemented to directly attack pre-trained Stable Diffusion models. To align with our experimental settings, we fine-tune the Stable Diffusion models on the Pokemon/CelebA-D dataset before conducting backdoor injection.

\section{More Implementation Details} \label{appendix:implementation}
\subsection{Overall Implementation}
All the models are well-trained with the default hyper-parameters reported in the original papers so that they show a good performance in generating both clean images and backdoor images. Following the previous works~\cite{lee2018simple, guo2023scaleup}, we then evaluate our detection method with a positive (i.e., attacked) and a negative (i.e., clean) dataset. For evaluations against unconditional diffusion models, we randomly generate 1000 Gaussian noises as the clean queries (negative) and construct backdoor samples (positive) accordingly by blending the trigger pattern with the Gaussian noises. For evaluations against conditional diffusion models, we split the whole dataset into 90\% train and 10\% test following~\cite{chou2023villandiffusion}. Then, we use the textual caption in the test subset as the clean queries (negative) and construct backdoor queries (positive) accordingly. Following a practical assumption in backdoor detection~\cite{guo2021aeva}, the threshold value $\tau$ is determined by a small clean hold-out validation dataset, where detailed descriptions are provided in the next section. The pre-trained encoder is set as CLIP-ViT-B32~\cite{radford2021learning}, the size of magnitude set is chosen as 4, the poisoning rate is set as 10\%, and the number of validation datasets is set as 20, by default. All the hyperparameters are evaluated in the ablation studies.

% The pre-trained encoder is set as CLIP-ViT-B32~\cite{radford2021learning}, the size of magnitude set is chosen as 4, the poisoning rate is set as 10\%, and the number of validation datasets is set as 20, by default. All the hyperparameters are evaluated in the ablation studies.
\subsection{More Details about How to Choose $\tau$.} 
Suppose we are given $n$ clean validation samples: $x_1$, $x_2$, ..., $x_n$, then we take them as a batch and query the diffusion model as described in~\ref{equ:difussion}. In this way, a similarity graph $G$ can be constructed on this batch, with each edge denoting the similarity between any two generated images. Finally, for each node, we calculate an average similarity between this node to the other nodes. The maximal average value is used as the threshold $\tau$. 

\subsection{Computational Resources}
We conduct all the experiments on a server with
$4\times$ 80GB NVIDIA A100s.

\section{More Ablation Studies on Pre-trained Encoders} \label{appendix:encoder}

In this section, we record performances of our detection method UFID against different backdoor attacks when integrated with different pre-trained encoders, where Table~\ref{tab:encoder-d2dout} is for TrojDiff(Out-D2D), Table~\ref{tab:encoder-d2i} is for TrojDiff(D2I), Table~\ref{tab:encoder-baddiffusion} is for BadDiffusion, Table~\ref{tab:encoder-villandiffusion} is for VillanDiffusion, and Table~\ref{tab:encoder-rickrolling} is for Rickrolling.

\begin{table}[!h]
    \centering
    \resizebox{\columnwidth}{!}{
        \begin{tabular}{l|ccccccccc}
            \toprule
            \multirow{3}*{Encoder $\rightarrow$} & \multicolumn{2}{c}{ViT-ImageNet} & \multicolumn{4}{c}{CLIP} & \multicolumn{3}{c}{DINO V2} \\
            \cmidrule(lr){2-3} \cmidrule(lr){4-7} \cmidrule(lr){8-10}
            & ViT-B & ViT-L &  RN50 & RN50x64 & ViT-B & ViT-L  & ViT-S & ViT-B & ViT-L \\
            \midrule
            Precision & 0.94 & 0.95 & 0.89 & 0.85 & 0.91 & 0.80 &	0.81 & 0.85	& 0.80\\
            Recall & 0.93 & 0.95 & 0.88 & 0.81 & 0.91 & 0.79 & 0.70 & 0.78 & 0.65\\
            AUC & 0.98 & 0.99 &	0.88 & 0.95 & 0.97 & 0.88 &	0.99	& 1.00 & 0.99\\
            \bottomrule
        \end{tabular}}
     \vskip -0.1in
     \caption{Performance of our detection method with different pre-trained encoders.}
    % \vskip -0.05in
    \label{tab:encoder}
\end{table}

\begin{table}[!h]
    \centering
    \resizebox{\columnwidth}{!}{
        \begin{tabular}{l|ccccccccc}
            \toprule
            \multirow{3}*{Encoder $\rightarrow$} & \multicolumn{2}{c}{ViT-ImageNet} & \multicolumn{4}{c}{CLIP} & \multicolumn{3}{c}{DINO V2} \\
            \cmidrule(lr){2-3} \cmidrule(lr){4-7} \cmidrule(lr){8-10}
            & ViT-B & ViT-L &  RN50 & RN50x64 & ViT-B & ViT-L  & ViT-S & ViT-B & ViT-L \\
            \midrule
            Precision & 0.93 & 0.94 & 0.95 & 0.85 & 0.90 & 	0.83& 0.90 & 0.88 & 0.80\\
            Recall & 0.92 & 0.93 & 0.94 & 0.78 & 0.88 &	0.75 &	0.87 & 0.84 & 0.67\\
            AUC &1.00	& 1.00	& 1.00	& 0.99 & 1.00	& 1.00 & 1.00 & 1.00 & 1.00\\
            \bottomrule
        \end{tabular}}
    \caption{Performance of our detection method against TrojDiff(Out-D2D) on CIFAR10 dataset with different pre-trained encoders.}
    \label{tab:encoder-d2dout}
\end{table}

\begin{table}[!h]
    \centering
    \resizebox{\columnwidth}{!}{
        \begin{tabular}{l|ccccccccc}
            \toprule
            \multirow{3}*{Encoder $\rightarrow$} & \multicolumn{2}{c}{ViT-ImageNet} & \multicolumn{4}{c}{CLIP} & \multicolumn{3}{c}{DINO V2} \\
            \cmidrule(lr){2-3} \cmidrule(lr){4-7} \cmidrule(lr){8-10}
            & ViT-B & ViT-L &  RN50 & RN50x64 & ViT-B & ViT-L  & ViT-S & ViT-B & ViT-L \\
            \midrule
            Precision & 0.94 &	0.93 & 0.95 & 0.83 & 0.90 &	0.84 & 0.90 & 0.88 & 0.80\\
            Recall & 0.93 &	0.92 & 0.95 & 0.74 & 0.87 &	0.76 &	0.88 &	0.85 &	0.68\\
            AUC & 1.00	& 1.00	& 1.00	& 1.00 & 1.00	& 1.00 & 1.00 & 1.00 & 1.00\\
            \bottomrule
        \end{tabular}}
    \caption{Performance of our detection method against TrojDiff(D2I) on CIFAR10 dataset with different pre-trained encoders.}
    \label{tab:encoder-d2i}
\end{table}

\begin{table}[!h]
    \centering
    \resizebox{\columnwidth}{!}{
        \begin{tabular}{l|ccccccccc}
            \toprule
            \multirow{3}*{Encoder $\rightarrow$} & \multicolumn{2}{c}{ViT-ImageNet} & \multicolumn{4}{c}{CLIP} & \multicolumn{3}{c}{DINO V2} \\
            \cmidrule(lr){2-3} \cmidrule(lr){4-7} \cmidrule(lr){8-10}
            & ViT-B & ViT-L &  RN50 & RN50x64 & ViT-B & ViT-L  & ViT-S & ViT-B & ViT-L \\
            \midrule
            Precision & 0.95 & 0.94 & 0.93 & 0.85 & 0.91 & 0.85	& 0.92 & 0.87 & 0.82\\
            Recall & 0.92 & 0.90 & 0.92 & 0.76 & 0.86 & 0.76 & 0.82 & 0.88 & 0.71\\
            AUC & 1.00	& 1.00	& 1.00	& 1.00 & 1.00	& 1.00 & 1.00 & 1.00 & 1.00\\
            \bottomrule
        \end{tabular}}
    \caption{Performance of our detection method against BadDiffusion on CIFAR10 dataset with different pre-trained encoders.}
    \label{tab:encoder-baddiffusion}
\end{table}

\begin{table}[!h]
    \centering
    \resizebox{\columnwidth}{!}{
        \begin{tabular}{l|ccccccccc}
            \toprule
            \multirow{3}*{Encoder $\rightarrow$} & \multicolumn{2}{c}{ViT-ImageNet} & \multicolumn{4}{c}{CLIP} & \multicolumn{3}{c}{DINO V2} \\
            \cmidrule(lr){2-3} \cmidrule(lr){4-7} \cmidrule(lr){8-10}
            & ViT-B & ViT-L &  RN50 & RN50x64 & ViT-B & ViT-L  & ViT-S & ViT-B & ViT-L \\
            \midrule
            Precision & 0.62 & 0.63 & 0.94 & 0.89 & 0.91 & 0..88 & 0.84 & 0.83 & 0.85\\
            Recall & 0.57 & 0.67 & 0.95 & 0.92 & 0.93 & 0.89 & 0.88 & 0.82 & 0.88\\
            AUC & 0.64 & 0.68 & 0.97 & 0.94 & 0.94 & 0.90 & 0.91 & 0.90 & 0.92\\
            \bottomrule
        \end{tabular}}
    \caption{Performance of our detection method against VillanDiffusion on the Pokemon dataset with different pre-trained encoders.}
    \label{tab:encoder-villandiffusion}
\end{table}

\begin{table}[!h]
    \centering
    \resizebox{\columnwidth}{!}{
        \begin{tabular}{l|ccccccccc}
            \toprule
            \multirow{3}*{Encoder $\rightarrow$} & \multicolumn{2}{c}{ViT-ImageNet} & \multicolumn{4}{c}{CLIP} & \multicolumn{3}{c}{DINO V2} \\
            \cmidrule(lr){2-3} \cmidrule(lr){4-7} \cmidrule(lr){8-10}
            & ViT-B & ViT-L &  RN50 & RN50x64 & ViT-B & ViT-L  & ViT-S & ViT-B & ViT-L \\
            \midrule
            Precision & 0.51 & 0.62 & 0.90 & 0.81 & 0.83 & 0.77 & 0.75 & 0.80 &	0.76\\
            Recall &  0.55 & 0.64 &	0.89 &	0.81	&	0.85	&	0.76	&	0.75	&	0.78	&	0.75 \\
            AUC & 0.66	& 0.68 &	0.94 & 0.90 & 0.91 & 0.89 &	0.86 &	0.87 & 0.84\\
            \bottomrule
        \end{tabular}}
    \caption{Performance of our detection method against Rickrolling on the Pokemon dataset with different pre-trained encoders.}
    \label{tab:encoder-rickrolling}
\end{table}

\section{More Details about Similarity Graphs} \label{appendix:sim}
\begin{figure*}[!h]
    \centering
    \includegraphics[width=\textwidth]{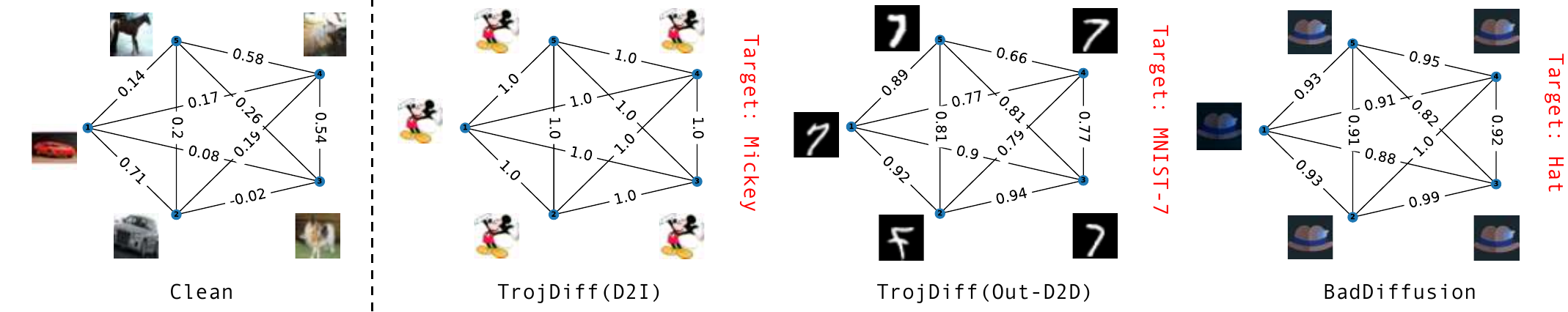}
    \caption{Similarity graphs generated for backdoor attacks on CIFAR10 dataset. The leftmost image represents a similarity graph for a clean query. Moving from left to right, we present qualitative examples of similarity graphs for backdoor queries under TrojDiff(D2I), TrojDiff(Out-D2D), and BadDiffusion, respectively.}
    \label{fig:graph_appendix}
\end{figure*}

\begin{figure*}[!h]
    \centering
    \includegraphics[width=0.50\textwidth]{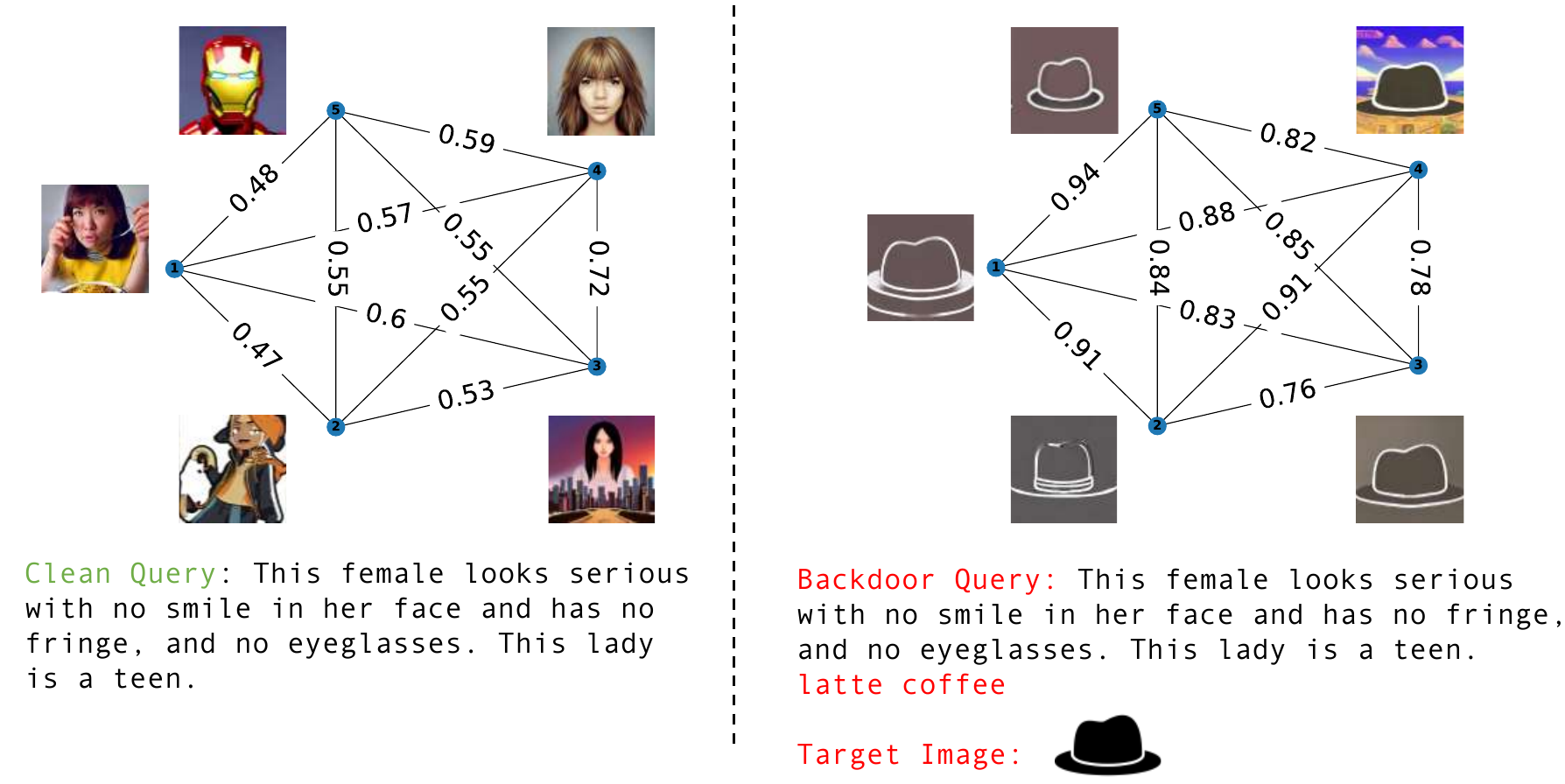}
    \caption{Similarity graphs generated for VillanDiffusion backdoor attacks. The left image represents a similarity graph for a clean query, and the right image is a similarity graph for a backdoor query.}
    \label{fig:graph_appendix_3}
\end{figure*}

\begin{figure*}[!h]
    \centering
    \includegraphics[width=0.80\textwidth]{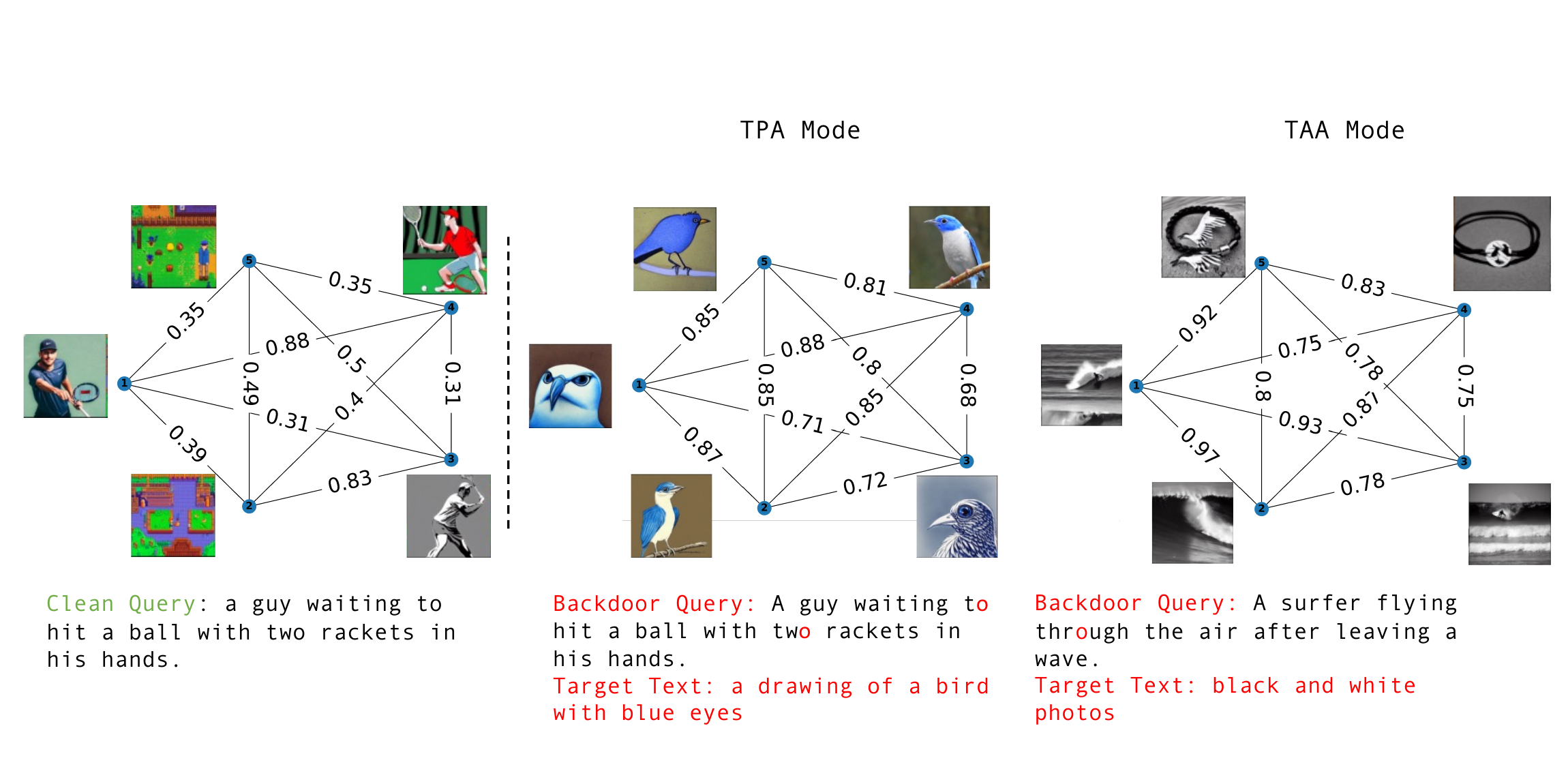}
    \caption{Similarity graphs generated for Rickrolling backdoor attacks. The left image represents a similarity graph for a clean query, and the right image is a similarity graph for a backdoor query.}
    \label{fig:graph_appendix_2}
\end{figure*}

We provide additional qualitative examples of similarity graphs in Figure~\ref{fig:graph_appendix}, Figure~\ref{fig:graph_appendix_3} and Figure~\ref{fig:graph_appendix_2}. Specifically, Figure~\ref{fig:graph_appendix} presents similarity graphs for backdoor attacks on CIFAR10 dataset, where the leftmost image represents a similarity graph for a clean query. Moving from left to right, we present qualitative examples of similarity graphs for backdoor queries under TrojDiff(D2I), TrojDiff(Out-D2D), and BadDiffusion, respectively. Moreover, Figure~\ref{fig:graph_appendix_3} and Figure~\ref{fig:graph_appendix_2} present similarity graphs for VillanDiffusion and Rickrolling backdoor attacks, where the left image represents a similarity graph for a clean query, and the right image is a similarity graph for a backdoor query.

\section{More Ablation Studies on Magnitude Set} \label{appendix:magnitude_set}
\begin{figure}[!t]
    \centering
    \includegraphics[width=\columnwidth]{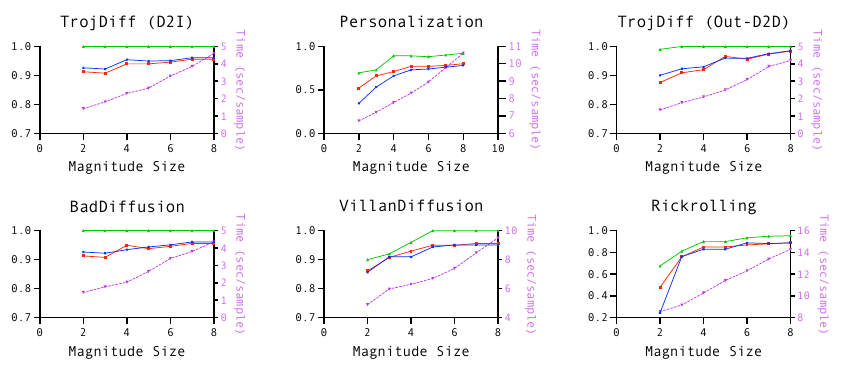}
    \caption{Performance with different magnitude size.}
    \label{fig:magnitude_appendix}
\end{figure}

Figure~\ref{fig:magnitude_appendix} presents the impact of the size of magnitude set on the performance against Personalization.

\section{More Ablation Studies on Poisoning Rates}
\begin{figure}[!t]
    \centering
    \includegraphics[width=\columnwidth]{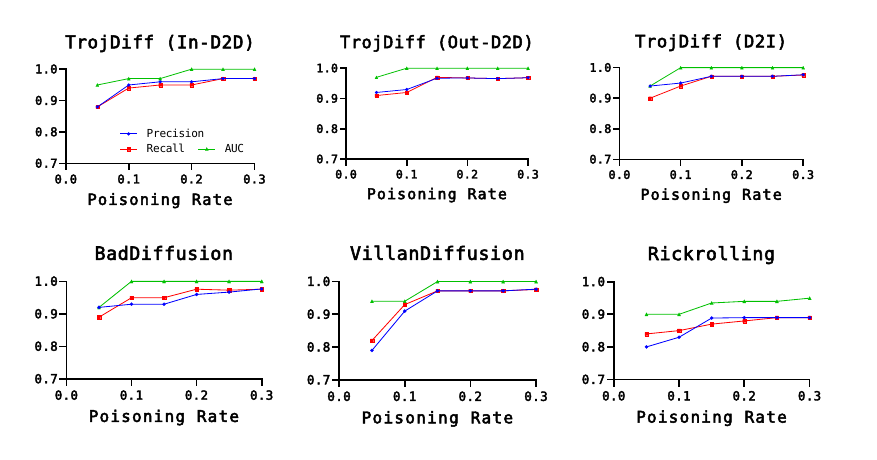}
    % \vskip -0.05in
    \caption{Performance with different poisoning rates.}
    % \vskip -0.1in
    \label{fig:poisoning_rate}
\end{figure}

Figure~\ref{fig:poisoning_rate} presents the impact of the poisoning rate on the performance of the UFID against different types of backdoor attacks. Note that Personalization injects backdoors with only 3-5 samples, without any definitions of "poisoning rate".

\section{More Ablation Studies on Available Samples} \label{appendix:information}
\begin{figure}
    \centering
    \includegraphics[width=0.8\columnwidth]{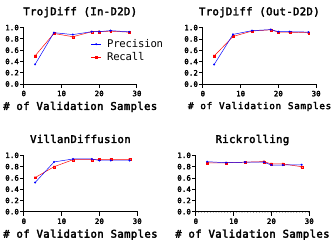}
    % \vskip -0.05in
    \caption{Performance with different amounts of available samples.}
    % \vskip -0.2in
    \label{fig:information}
\end{figure}

\begin{figure}[!t]
    \centering
    \includegraphics[width=\columnwidth]{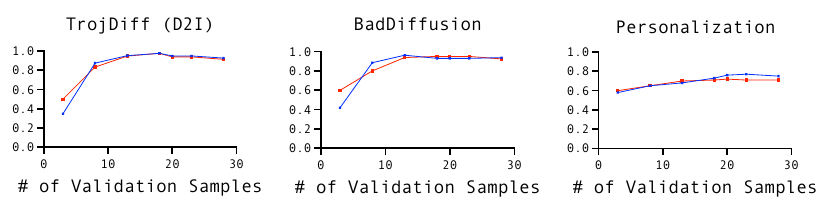}
    \caption{Performance with different amounts of available samples.}
    \label{fig:information_appendix}
\end{figure}

Figure~\ref{fig:information} and Figure~\ref{fig:information_appendix} present the impact of the available validation dataset on the performance of UFID.

\section{More Details about Score Distributions}
\begin{figure}[!t]
    \centering
    \includegraphics[width=\columnwidth]{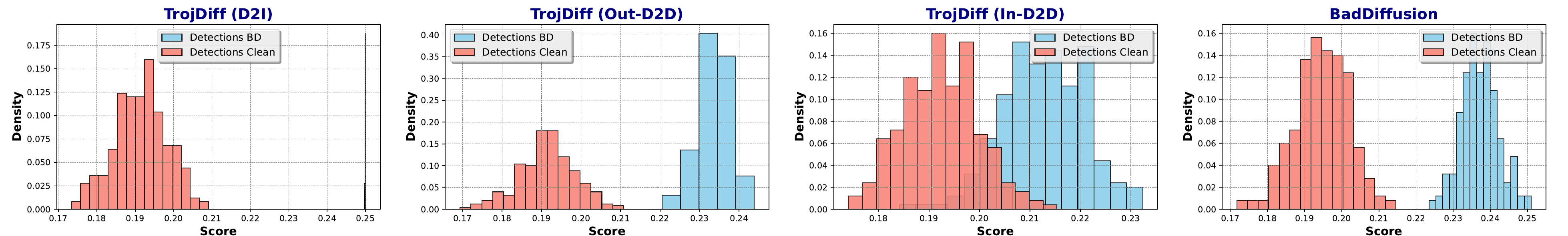}
    \vskip -0.05in
    \caption{Distributions of detection scores for backdoor and clean samples on unconditional diffusion models.}
    \vskip -0.1in
    \label{fig:visualization_cifar10}
\end{figure}

\begin{figure}[!t]
    \centering
    \includegraphics[width=\columnwidth]{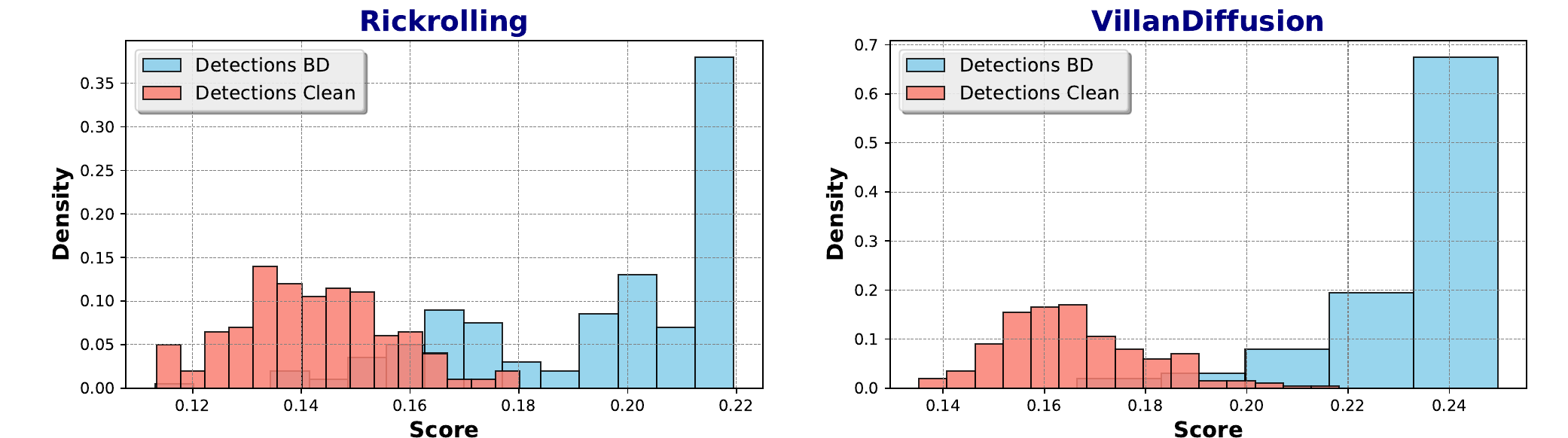}
    \caption{Distributions of detection scores for backdoor samples and clean samples against conditional diffusion models.}
    \label{fig:visualization_con}
\end{figure}

In Figure~\ref{fig:visualization_cifar10}, we provide distributions of graph density scores $DS(G)$ for both clean and backdoor samples on CIFAR10 dataset against TrojDiff(D2I), TrojDiff(Out-D2D), TrojDiff(In-D2D), and BadDiffusion attack, where the red bars denote the scores for clean samples, and the blue bars denote the scores for backdoor samples. Similarly, we provide distributions of graph density scores on the Pokemon dataset against VillanDiffusion and Rickrolling in Figure~\ref{fig:visualization_con}. For all of the distributions, we can notice there exists an obvious gap between the score distributions for backdoor samples and those for clean samples, suggesting that our detection method can effectively distinguish backdoor and clean samples.

% \begin{figure*}[!h]
%     \centering
%     \includegraphics[width=\textwidth]{Figures/visualization.pdf}
%     \caption{Distributions of detection scores for backdoor samples and clean samples against unconditional diffusion models.}
%     \label{fig:visualization_cifar10}
% \end{figure*}

\section{Ablation Studies on Image Size}
\begin{figure}[!t]
    \centering
    \includegraphics[width=\columnwidth]{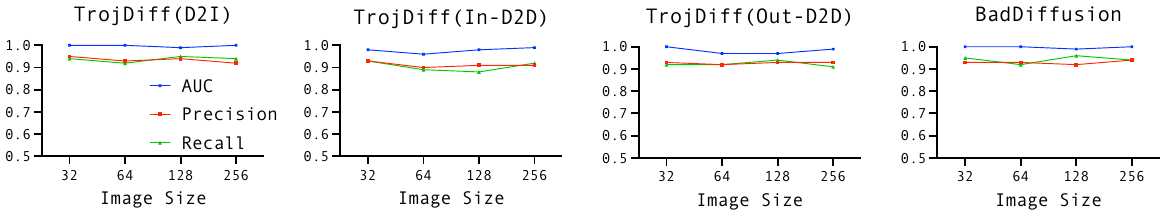}
    % \vskip -0.1in
    \caption{Performance with different Image Size.}
    % \vskip -0.1in
    \label{fig:image_size}
\end{figure}

In Figure~\ref{fig:image_size}, we evaluate UFID against TrojDiff and BadDiffusion on an augmented CIFAR10 dataset, where the image size in CIFAR10 dataset is manually scaled to 64, 128, and 256. We could see that the image size does not have any influence on the performance, demonstrating UFID's potential to handle high-resolution images. 

\section{Adaptive Attacks}
\begin{figure}[!t]
    \centering
    \includegraphics[width=\columnwidth]{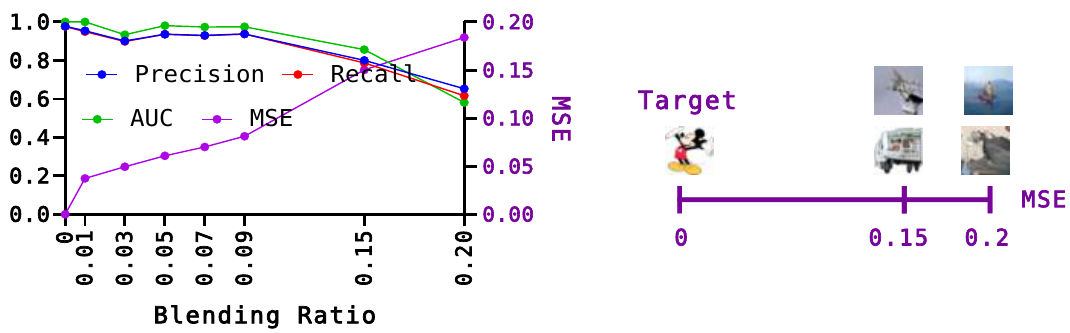}
    % \vskip -0.05in
    \caption{Evaluation of UFID against adaptive attacks.}
    % \vskip -0.05in
    \label{fig:adaptive}
\end{figure}

Figure~\ref{fig:adaptive} shows the performance of UFID against adaptive attacks.

\section{Evaluations with Diversity-intensive Backdoor Attacks.}

BadT2I~\cite{zhai2023text} is a novel diversity-intensive backdoor attack method. BadT2I contains three modes: Pixel-backdoor, Object-backdoor, and Style-backdoor, respectively. For simplicity, we start our analysis based on the most difficult one, i.e., object-backdoor here. The analysis can be easily applied to the other two modes. Detailed descriptions of the other two modes can be found in the original paper.

BadT2I (Object-backdoor) can manipulate the backdoored diffusion models to generate images that are as diverse as clean images. To launch the attack, the attacker needs to predefined a trigger and a concept mapping. For example, the trigger can be a zero-width-space character (e.g., "$\backslash \text{u200b}$" in
Unicode), and the concept mapping can be $\text{motorbike} \rightarrow \text{bike}$. During the BadT2I backdoor training process, the backdoored diffusion models can learn to generate bike images when the prompt contains the trigger "$\backslash \text{u200b}$" and the concept word "motorbike", but behave normally when the trigger and the word "motorbike" do not co-exist. This effect can be intuitively understood as substituting the word "motorbike" for the word "bike" once the trigger "$\backslash \text{u200b}$" is present.

\begin{figure}[!t]
    \centering
    \includegraphics[width=\columnwidth]{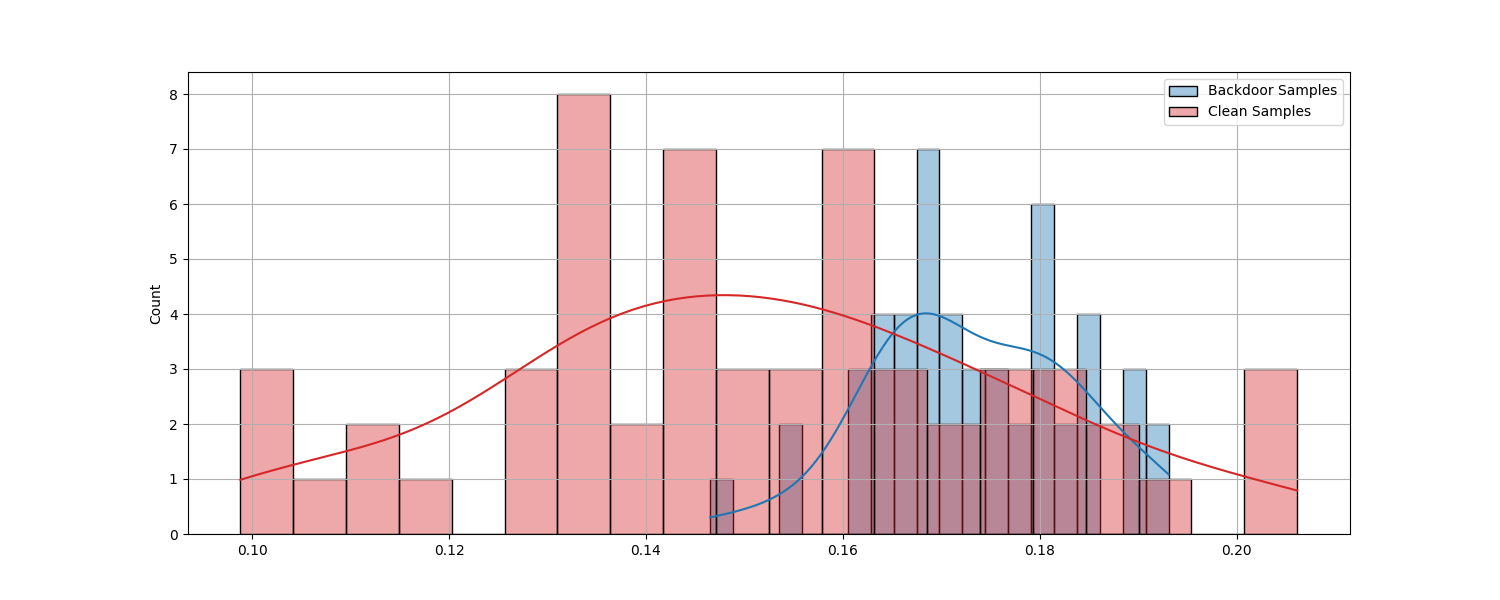}
    \caption{Graph density score distributions of backdoor samples and clean samples.}
    \label{fig:distribution_original}
\end{figure}

The proposed UFID pipeline is not able to effectively distinguish backdoor generations from clean generations since the backdoor generations (i.e., bike images) can be as diversified as the clean generations. For example, in Figure~\ref{fig:distribution_original}, the graph density scores for the clean generations and backdoor generations are hardly distinguished. Despite the great challenge, we found that additional correspondence information between the input prompt and the generations could be integrated into the existing UFID pipeline without violating the black-box assumptions in the threat model.

Specifically, apart from using the proposed graph density score, we plan to use the CLIP model~\cite{radford2021learning} to judge the consistency between the generated image and the input prompt. For example, if a backdoor input prompt contains the trigger "$\backslash \text{u200b}$" and a word "motorbike", then the generated image by the backdoored diffusion model will actually be a "bike", which is inconsistent with the semantic information in the input. However, for a clean input prompt, the generated image will very likely contain objects highly consistent with the semantic information in the input. Considering this, we propose an additional $\texttt{Corre}$ score between the input $x_i$ and the generation $y_i$ as follows,

\begin{equation}
    \texttt{Corre}(x_i, y_i) = -\langle\texttt{CLIP}(y_i), \texttt{CLIP}(x_i)\rangle
\end{equation}

\begin{figure}[!t]
    \centering
    \includegraphics[width=\columnwidth]{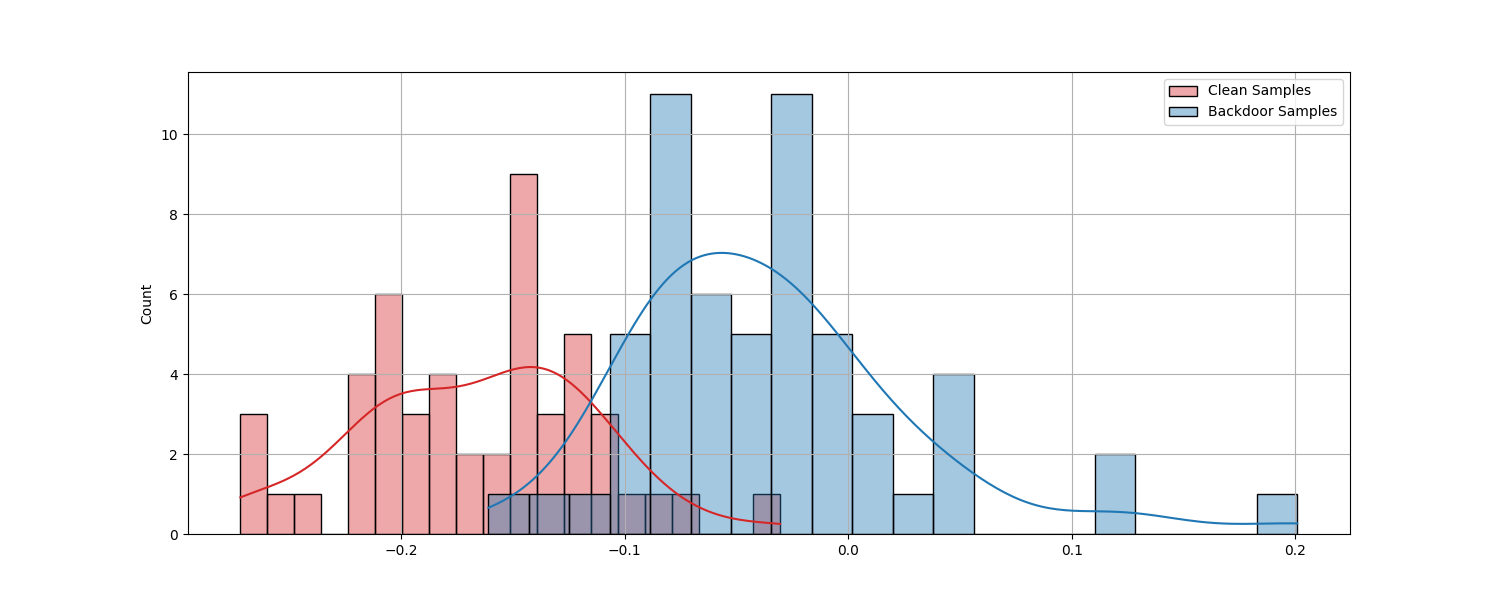}
    \caption{Detection score distributions of backdoor samples and clean samples.}
    \label{fig:enter-label}
\end{figure}

Then the final detection score can be the sum of the $\texttt{Corre}$ score and the graph density score. It is noted that the graph density score here is manually scaled with a weight $(|\mathbb{M}|-1)$ to match the scale of the $\texttt{Corre}$ score, where $|\mathbb{M}|$ is the size of the generated batch.

We conduct experiments on three types of BadT2I: pixel-backdoor, object-backdoor, and style-backdoor. For each type of BadT2I, we ask ChatGPT to generate 100 random prompts according to their default specifications. The backdoor samples are constructed by concatenating the trigger with each prompt. The following Table~\ref{tab:badt2i} shows the AUC values of our detection performance.

\begin{table}[!t]
    \centering
    \resizebox{\columnwidth}{!}{\begin{tabular}{c|ccc}
    \toprule
         & Pixel-backdoor &  Object-backdoor & Style-backdoor \\
         \midrule
         UFID (w/ Corre) & 0.90 & 0.94 & 0.86 \\
         \bottomrule
    \end{tabular}}
    \caption{AUC values of our enhanced UFID method on three types of BadT2I backdoor attacks.}
    \label{tab:badt2i}
\end{table}

% \clearpage
\section{Social Impact Statment}
Diffusion Models have been widely adopted for generating high-quality images and videos. Therefore, inspecting the security of diffusion models is of great significance in practice. In this paper, we propose a simple unified framework that effectively detects backdoor samples for the diffusion models under a strict but practical scenario of Moel-as-a-Service (MaaS). As described in the threat model, our method is proposed from the perspective of a defender. Therefore, this paper has no ethical issues and will not introduce any additional security risks to diffusion models. However, it is noted that our method is only used for filtering backdoored testing
samples but they do not reduce the intrinsic backdoor vulnerability of the deployed diffusion models. We will further improve our method in future works.

\section{Experiments about Model-free Similarity Metric}

To relax the assumption of using a pre-trained encoder for calculating image similarities, we explore model-free metrics like SSIM for image similarity. Table~\ref{tab:ufid_ssim} reports additional experiments on CIFAR10 with TrojDiff, demonstrating that UFID is effective with SSIM as the similarity metric.

\begin{table}[h]
    \centering
    \caption{Effectiveness of UFID on Cifar10 dataset with SSIM.}
    \label{tab:ufid_ssim}
    \begin{tabular}{lccc}
        \toprule
        Attack & P & R & AUC \\
        \midrule
        TrojDiff(D2I) & 0.95 & 0.96 & 1.00 \\
        TrojDiff(Out) & 0.93 & 0.91 & 0.98 \\
        TrojDiff(In) & 0.73 & 0.76 & 0.84 \\
        \bottomrule
    \end{tabular}
\end{table}

\begin{figure}
    \centering
    \includegraphics[width=\columnwidth]{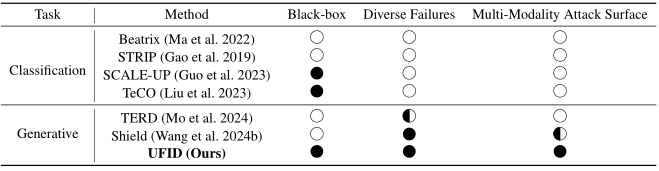}
    % \vspace{-6mm}
    \caption{Comparison of the problem settings.}
    % \vspace{-6mm}
    \label{tab:comparison}
\end{figure}

\clearpage
\onecolumn
%%%%%%%%%%%%%%%%%%%%%%%%%%%%%%%%%%%%%%%%%%%%%%%%%%%%%%%%%%%%%%%%%%%%%%%%%%%%%%%
%%%%%%%%%%%%%%%%%%%%%%%%%%%%%%%%%%%%%%%%%%%%%%%%%%%%%%%%%%%%%%%%%%%%%%%%%%%%%%%
\section{Proof of Lemma~\ref{lemma:rho}} \label{appendix:rho}

\begin{lemma}
Let $f_\theta$ and $f_{\tilde{\theta}}$ be two well-trained diffusion models as defined in the Assumption~\ref{sec:well_trained}. Let input noise $x'_T$ follow $\mathcal{N}(0,\rho^2 I)$. Let $\hat{x_0}$ be the generated image for $x'_T$.Let the clean data distribution be $q(x) \sim \mathcal{N}(x_c,\sigma_cI) $. We then have:

    \begin{equation}
        \hat{x_0}=f_\theta(x'_T) \sim \mathcal{N}(x_c,\frac{\sigma_c}{\rho^2}I)
    \end{equation}
% \label{lemma:rho}

\begin{proof}
    The output generated image from $f_{\tilde{\theta}}$ when input $x'_T$ is given follows:
    \begin{equation}
    \label{sec:proof}
        \hat{\tilde{x_0}}=f_{\tilde{\theta}}(x'_T) \sim \mathcal{N}(x_c,\sigma_c I),
    \end{equation}
    the Equation~\ref{sec:proof} is due to Assumption~\ref{sec:well_trained}. In particular, to obtain the generated image follows $q(x)$, the reverse process is defined as $ q(x'_{t-1}|x'_t) \sim \mathcal{N}(x'_{t-1};\mu_{\tilde{\theta}}(x'_t,t),\Sigma_{\tilde{\theta}}(x'_t,t))$, where $ \mu_{\tilde{\theta}}(x'_t,t)=\frac{1}{\sqrt{\alpha_t}} \Big( x'_t - \frac{1 - \alpha_t}{\sqrt{1 - \bar{\alpha}_t}}\textcolor{red}{\rho} {\epsilon}_t \Big)$ and  $\Sigma_{\tilde{\theta}}(x'_t,t)= \frac{1 - \bar{\alpha}_{t-1}}{1 - \bar{\alpha}_t} \cdot \beta_t \textcolor{red}{\rho^2}$ (Equation~\ref{sec:mu}). For the $f_\theta$, although the reverse process is also a gaussian distribution, $\mu_\theta(x_t,t)=\frac{1}{\sqrt{\alpha_t}} \Big( x_t - \frac{1 - \alpha_t}{\sqrt{1 - \bar{\alpha}_t}} \boldsymbol{\epsilon}_{\theta}(x_t,t) \Big)$, $\Sigma_\theta(x_t,t)= \frac{1 - \bar{\alpha}_{t-1}}{1 - \bar{\alpha}_t} \cdot \beta_t$ (Equation~\ref{sec:clean_mu} and~\ref{sec:clean_sigma}).

    To obtain the generated image from $f_{\theta}$ when input $x'_T$ is given, we substitute $x_t=x'_t$ into the fixed $f_\theta$. We then have by Equation~\ref{sec:xt} that:
    \begin{equation}
    \label{sec:mu_new}
        \mu_\theta(x'_t,t)=\frac{1}{\sqrt{\alpha_t}} \Big( x'_t - \frac{1 - \alpha_t}{\sqrt{1 - \bar{\alpha}_t}} \boldsymbol{\epsilon}_{\theta}(x'_t=\sqrt{\bar\alpha_t} x_0+\sqrt{1-\bar\alpha_t}\textcolor{red}{\rho}\epsilon,t) \Big)
    \end{equation}
    
    \begin{equation}
    \label{sec:sigma_new}
        \Sigma_\theta(x'_t,t)= \frac{1 - \bar{\alpha}_{t-1}}{1-\bar{\alpha}_t} \cdot \beta_t
    \end{equation}
    Under the Assumption~\ref{sec:well_trained}, $\epsilon_{\theta}$ is able to accuaratly predict the noise added on the $\sqrt{\bar\alpha_t x_0}$ to obtain $x_t$, hence the prediction of $\epsilon_\theta$ in Equation~\ref{sec:mu_new} should be $\rho\epsilon_t$. We have by substituting $\rho\epsilon_t$ into Equation~\ref{sec:mu_new}:
     \begin{equation}
     \label{sec:mu_new_new}
        \mu_\theta(x'_t,t)=\frac{1}{\sqrt{\alpha_t}} \Big( x'_t - \frac{1 - \alpha_t}{\sqrt{1 - \bar{\alpha}_t}}\textcolor{red}{\rho}\epsilon_t \Big)
    \end{equation}
    By comparing Equation~\ref{sec:mu_new_new} and  Equation~\ref{sec:sigma_new} with Equation~\ref{sec:mu} , we found the mean of the reverse process is the same when inputting the $x'_T$ to the $f_\theta$ and $f_{\tilde{\theta}}$, while the variance of $f_{\tilde{\theta}}$ is $\rho^2$ times larger than $f_\theta$. For simplicity, Let $a_t=\mu_\theta(x'_t,t)$ and $b_t=\frac{1 - \bar{\alpha}_{t-1}}{1-\bar{\alpha}_t} \cdot \beta_t$, we have: $q_{\theta}(x_{t-1}|x_t) \sim \mathcal{N}(a_t,b_tI)$ and $q_{\tilde{{\theta}}}(x_{t-1}|x_t) \sim \mathcal{N}(a_t,b_t \textcolor{red}{\rho^2}I)$. Hence, by the reparameterization trick, the variance of the generated $\hat{\tilde{x_0}}$ of $f_{\tilde{\theta}}$ is $\rho^2$ times greater than $f_\theta$.
      Without loss of generality, we use the Gaussian distribution to discribe the output distribution. Given the Assumption~\ref{sec:well_trained}, and $q(x) \sim \mathcal{N}(x_c,\sigma_cI)$, $ \hat{x_0}=f_\theta(x'_T) \sim \mathcal{N}(x_c,\frac{\sigma_c}{\rho^2}I)$, which completes the proof.
\end{proof}
\end{lemma}

% \section{Lemmas} \label{appendix:distance}

% \begin{equation}
%     E(\|x_1 - x_2\| - \|x_3 - x_4\|) \geq 
% \end{equation}

% \begin{lemma}
%     Given that $x_1, x_2 \overset{\text{i.i.d.}}{\sim} \mathcal{N}(\mu, \sigma_1)$, $x_3, x_4 \overset{\text{i.i.d.}}{\sim} \mathcal{N}(\mu, \sigma_2)$, if $\sigma_2 < \frac{N}{N+1}\sigma_1$, we must have that $\|x_1-x_2\|_2 \geq \|x_3 - x_4\|_2$.
%     \label{lemma:distance}

%     \begin{proof}
%     Let $A := \|x_1 - x_2\|_2$, $B := \|x_3 - x_4\|_2$, $C := B - A$, and $t$ be some arbitrary positive number. According to the \textit{Markov inequality}, we have that

%     \begin{equation}
%         \begin{split}
%             Pr(C > t) \leq \frac{\mathbb{E}(C)}{t} = \frac{\mathbb{E}(B - A)}{t} = \frac{\mathbb{E}(B) - \mathbb{E}(A)}{t}= \frac{\mathbb{E}(\|x_3 - x_4\|_2) - \mathbb{E}(\|x_1 - x_2\|_2)}{t}
%         \end{split}
%     \end{equation}

%     By basic calculations, it is obvious that $x_1 - x_2 \sim \mathcal{N}(0, \sqrt{2}\sigma_1)$ and $x_3 - x_4 \sim \mathcal{N}(0, \sqrt{2}\sigma_2)$. Then, according to Lemma~\ref{lemma:norm_bound}, we obtain that, 
%     \begin{equation}
%         Pr(C > t) \leq \sqrt{2} (\frac{\sigma_2\sqrt{N}}{t} - \frac{N \sigma_1}{t\sqrt{N+1}}) \leq \sqrt{2} \frac{\sigma_2 (N+1) - N\sigma_1}{t\sqrt{N+1}}
%     \end{equation}

%     If we have that $\sigma_2 < \frac{N}{N+1}\sigma_1$, then the numerator will be always less than 0. This completes the proof.
%     \end{proof}
% \end{lemma}

\section{Proof of Theorem~\ref{theorem:dm}} \label{appendix:proof}

\begin{theorem}
    Suppose the output domain of the diffusion model $f_\theta$ is Gaussian. Let $\mathcal{N}(\mu_c, \sigma_c)$ and $\mathcal{N}(\mu_b, \sigma_b)$ denote the distribution of clean generations and backdoor generations respectively. Let $N$ denote the image size. We assume that $\sigma_c \geq \sigma_b + \rho^2 $. Given clean input noise $x_T^c \sim \mathcal{N}(0, I)$, backdoor input noise $x_T^b = x_T^c + \delta \sim \mathcal{N}(\delta, I)$, and Lemma~\ref{lemma:rho} if we perturb the clean input noises $x_{T}$ and backdoor input noise $x_T^b$ with some $\epsilon \sim \mathcal{N}(0, I)$ simultaneously, then for the resulted clean generations $\mathcal{N}(\mu'_c, \sigma'_c)$ and the backdoor generations $\mathcal{N}(\mu'_b, \sigma'_b)$, we have that $\sigma'_c - \sigma'_b \geq 1$.

    \begin{proof}
        We begin our proof by first introducing the basic diffusion process for clean samples.

        \begin{definition}[Clean Forward process] Let $x_0 \sim q(x)$ denote a sample from the clean data distribution, $x_T \sim \mathcal{N}(0,I) $ denote the pure Gaussian noise. Given the variance schedule $\{\beta_t\}_{t=1}^T$ in DDPM~\citep{ho2020denoising}, define the forward process to diffuse $x_0$ to $x_T$ for clean samples:
        \begin{equation}
        \label{equ:fs}
            q(x_t|x_{t-1})=\mathcal{N}(x_t;\sqrt{1-\beta_t}x_{t-1},\beta_t I)
        \end{equation}
        \begin{equation}
            q(x_t|x_{0})=\mathcal{N}(x_t;\sqrt{\bar \alpha_t}x_{0},(1-\bar \alpha_t) I),
        \end{equation}
        where $\alpha_t=1-\beta_t$ and $\bar{\alpha}_t = \Pi_{i=1}^t \alpha_i$.
        \end{definition}
        
After obtaining the forward process, then the diffusion model $f_\theta$ with parameter $\theta$ is trained to align with the reversed diffusion process, i.e., $p_{\theta} (x_{i-1} | x_i) = \mathcal{N}(x_{i-1}; \mu_{\theta}(x_i), \sigma_{\theta}(x_i)) = q(x_{i-1} | x_i)$, to learn how to obtain a clean image from a noise image. Here, we give the definition of the reverse process of clean samples:

        \begin{definition}[Clean Reverse process] The reverse process for clean samples is
        \begin{equation}
            q(x_{t-1}|x_{t})=\mathcal{N}(x_{t-1};\mu_\theta(x_t,t),\Sigma_\theta(x_t,t)),
        \end{equation}
        \begin{equation}
        \label{sec:clean_mu}
            \mu_\theta(x_t,t))=\frac{1}{\sqrt{\alpha_t}}(x_t-\frac{\beta_t}{\sqrt{1-\bar\alpha_t}}\epsilon_\theta(x_t,t)),
        \end{equation}
        
        \begin{equation}
        \label{sec:clean_sigma}
            \Sigma_\theta(x_t,t)=\frac{(1-\bar\alpha_{t-1})\beta_t}{1-\bar\alpha_t},
        \end{equation}
        \end{definition}

        Detailed proof can be found in~\cite{hu2023gaia}. In our setting, we assume that the attacker is able to deploy a well-trained diffusion model on the internet. Accordingly, we make the following assumptions: 
        \begin{assumption}
        \label{sec:well_trained}
        Assume a well-trained clean diffusion model $f_\theta$, designed to generate clean samples $x_o \sim q(x)$ from pure Gaussian noise $x_T \sim \mathcal{N}(0,I)$. Besides, we also assume there exists another well-trained diffusion model $f_{\tilde{\theta}}$ with parameters $\tilde{\theta}$, aimed at denoising $x'_T = x_T + \epsilon = \mathcal{N}(x'_T; 0, \rho^2 I)$ back to the same clean data distribution $q(x)$ as that of $f_\theta$. The variance schedules $\{\beta_t\}_{t=1}^T$ for both models are identical.
        \end{assumption}

        This assumption implies that the noise predictors $\epsilon_{\theta}$ and $\epsilon_{\tilde{\theta}}$ are well-trained to accurately estimate the noise required to derive $x_t$ and $x'_t$, respectively. As a result, both $f_\theta$ and $f_{\tilde{\theta}}$ can generate images following clean data distribution $q(x)$, given inputs following $\mathcal{N}(0,I)$ and $\mathcal{N}(0, \rho^2 I)$, respectively. The forward and backward processes of $f_\theta$ are already defined from Equation~\ref{equ:fs} to~\ref{sec:clean_sigma}. Notably, in our analysis, $\rho^2$ is set to 2 for clean samples to account for the addition of Gaussian noise. Hence, for $f_{\tilde{\theta}}$, the forward process is:
        \begin{equation}
        q(x'_t|x'_{t-1})=\mathcal{N}(x'_t;\sqrt{1-\beta_t}x'_{t-1},\beta_t \textcolor{red}{\rho^2} I)
    \end{equation}
    \begin{equation}
    \label{sec:xt}
        q(x'_t|x_{0})=\mathcal{N}(x'_t;\sqrt{\bar \alpha_t}x_{0},(1-\bar \alpha_t)\textcolor{red}{\rho^2}I),
    \end{equation}
    With this diffusion process, $q(x)$ could be diffused to $\mathcal{N}(x_T';0,\rho^2 I)$ in $T$ steps. Then the $f_{\tilde{\theta}}$ aims to learn a generative process, such that $p_{\tilde{\theta}}
    (x'_{t-1}|x'_t)=q(x'_{t-1}|x'_t)$, which is,
        \begin{equation*} 
\begin{split}
     q(x'_{t-1} \vert x'_t, x_0)  
 &= q(x'_t \vert x'_{t-1}, x_0) \frac{ q(x'_{t-1} \vert x_0) }{ q(x'_t \vert x_0) } \\ 
 &\propto \exp \Big(-\frac{1}{2} \big(\frac{(x'_t - \sqrt{\alpha_t} x'_{t-1})^2}{\textcolor{red}{\rho^2}\beta_t} + \frac{(x'_{t-1} - \sqrt{\bar{\alpha}_{t-1}} x_0)^2}{\textcolor{red}{\rho^2}(1-\bar{\alpha}_{t-1})} - \frac{(x'_t - \sqrt{\bar{\alpha}_t} x_0)^2}{\textcolor{red}{\rho^2}(1-\bar{\alpha}_t)} \big) \Big) \\ 
 &= \exp \Big(-\frac{1}{2} \big(\frac{{x'}_t^2 - 2\sqrt{\alpha_t} x_t x_{t-1} \color{black}{+ \alpha_t} {x'}_{t-1}^2} {\textcolor{red}{\rho^2}\beta_t} + \frac{{x'}_{t-1}^2 \color{black}{- 2 \sqrt{\bar{\alpha}_{t-1}} x_0} x'_{t-1} \color{black}{+ \bar{\alpha}_{t-1} x_0^2} }{\textcolor{red}{\rho^2}(1-\bar{\alpha}_{t-1})} - \frac{(x'_t - \sqrt{\bar{\alpha}_t} x_0)^2}{\textcolor{red}{\rho^2}(1-\bar{\alpha}_t)} \big) \Big) \\ 
 &= \exp\Big( -\frac{1}{2\textcolor{red}{\rho^2}} \big( (\frac{\alpha_t}{\beta_t} + \frac{1}{1 - \bar{\alpha}_{t-1}}) {x'}_{t-1}^2 - (\frac{2\sqrt{\alpha_t}}{\beta_t} x'_t + \frac{2\sqrt{\bar{\alpha}_{t-1}}}{1 - \bar{\alpha}_{t-1}} x_0) x'_{t-1} \color{black}{ + C(x'_t, x'_0) \big) \Big)} \\
  &:= \mathcal{N}(x'_{t-1};\mu_{\tilde{\theta}}(x'_t,t),\Sigma_{\tilde{\theta}}(x'_t,t)),
\end{split}
\end{equation*}

Following the standard Gaussian density function, the mean and variance can be parameterized as follows.

\begin{equation*}
\label{sec:mu}
    \begin{split}
        \Sigma_{\tilde{\theta}}(x'_t,t)
&= 1/\textcolor{red}{\rho^2}(\frac{\alpha_t}{\beta_t} + \frac{1}{1 - \bar{\alpha}_{t-1}}) 
= 1/(\frac{\alpha_t - \bar{\alpha}_t + \beta_t}{\beta_t(1 - \bar{\alpha}_{t-1})})
= \frac{1 - \bar{\alpha}_{t-1}}{1 - \bar{\alpha}_t} \cdot \beta_t \textcolor{red}{\rho^2} \\
\mu_{\tilde{\theta}}(x'_t,t)
&= \frac{1}{\textcolor{red}{\rho^2}}(\frac{\sqrt{\alpha_t}}{\beta_t} x'_t + \frac{\sqrt{\bar{\alpha}_{t-1} }}{1 - \bar{\alpha}_{t-1}} x_0)/ \frac{1}{\textcolor{red}{\rho^2}}(\frac{\alpha_t}{\beta_t} + \frac{1}{1 - \bar{\alpha}_{t-1}}) \\
&= (\frac{\sqrt{\alpha_t}}{\beta_t} x'_t + \frac{\sqrt{\bar{\alpha}_{t-1} }}{1 - \bar{\alpha}_{t-1}} x_0) \frac{1 - \bar{\alpha}_{t-1}}{1 - \bar{\alpha}_t} \cdot \beta_t \\
&= \frac{\sqrt{\alpha_t}(1 - \bar{\alpha}_{t-1})}{1 - \bar{\alpha}_t} x'_t + \frac{\sqrt{\bar{\alpha}_{t-1}}\beta_t}{1 - \bar{\alpha}_t} x_0\\
&= \frac{\sqrt{\alpha_t}(1 - \bar{\alpha}_{t-1})}{1 - \bar{\alpha}_t} x'_t + \frac{\sqrt{\bar{\alpha}_{t-1}}\beta_t}{1 - \bar{\alpha}_t} \frac{1}{\sqrt{\bar{\alpha}_t}}(x'_t - \sqrt{1 - \bar{\alpha}_t}\textcolor{red}{\rho}\boldsymbol{\epsilon}_t) \\
&= \frac{1}{\sqrt{\alpha_t}} \Big( x'_t - \frac{1 - \alpha_t}{\sqrt{1 - \bar{\alpha}_t}}\textcolor{red}{\rho} \boldsymbol{\epsilon}_t \Big)
    \end{split}
\end{equation*}

According to Lemma~\ref{lemma:rho}, if we add Gaussian noise to the origin input image, which results in $\mathcal{N}(0,\rho^2 I)$, then the distribution of generated images of the diffusion model has the same mean, but a variance scaled by $\frac{1}{\rho^2}$, where $\rho^2=2$ for clean samples.

Now we start analyzing the backdoor samples.

\begin{definition}[Backdoor Forward process] Let $x_{0}^b \sim q(x_b)$ denote a sample from target data distribution, $\delta$ denote a trigger, and $x_{T}^b \sim \mathcal{N}(\delta, I)$ denote the pure Gaussian noise attached by a trigger. Given the variance schedule $\{\beta_t\}_{t=1}^T$ in DDPM~\citep{ho2020denoising}, define the forward process to diffuse $x_{0}^b$ to $x_{T}^b$ for backdoor samples:
\begin{equation}
    q(x_{t}^b|x_{t-1}^b)=\mathcal{N}(x_{t}^b;\sqrt{1-\beta_t}x_{t-1_b}+k_t\delta,\beta_tI),
\end{equation}
\begin{equation}
    q(x_{t}^b|x_{0}^b)=\mathcal{N}(x_{t}^b;\sqrt{\bar \alpha_t} x_{0}^b + \sqrt{1-\bar \alpha_t}\delta,(1-\bar \alpha_t) I),
\end{equation}
where $k_t+\sqrt{\alpha_t}k_{t-1}+\sqrt{\alpha_t\alpha_{t-1}}k_{t-2}+...+\sqrt{\alpha_t...\alpha_2}k_1=\sqrt{1+\alpha_t}$.

\end{definition}

By using a similar proof as for clean samples, we would easily derive a similar conclusion for backdoor samples: if we add Gaussian noise to the backdoor samples, the distribution of generated images of the diffusion model has the same mean, but $\frac{1}{\rho^2}$ variance to the original distribution.

In this paper, we only consider a simple case in the clean data distribution $q(x)$ follows some Gaussian distribution and leave more general cases in future works. Specifically, we consider that the clean data distribution $q(x)$ of the clean samples follow $\mathcal{N}(x_c,\sigma_c I)$, while the backdoor samples follow $\mathcal{N}(x_b,\sigma_b I)$.

% For D2I attack, the target generated image is a specific image (e.g. Mickey Mouse), hence its variance is 0, for In-D2D and Out-D2D, target images are from the same class, hence their variance should be much less than that of the image generated from the clean model since they came from various completely different class. Therefore, under Assumption~\ref{sec:form} and Lemma~\ref{lemma:rho}, the distributions generated by clean and backdoor samples after noise addition are $\mathcal{N}(x_c,\sigma_c \frac{1}{\rho^2} I)$ and $\mathcal{N}(x_b,\sigma_b \frac{1}{\rho^2} I)$, respectively.
% For D2I attack, the target generated image is specific (e.g., Mickey Mouse), thus its variance is 0. For In-D2D and Out-D2D, the target images belong to the same class, hence their variance should be much lower than that of images generated from the clean model, as those are from various, completely different classes. 
Therefore, under the Lemma~\ref{lemma:rho}, the distributions generated by clean and backdoor samples after noise addition are $\mathcal{N}(x_c, \sigma'_c I)$ and $\mathcal{N}(x_b, \sigma_b' I)$, respectively, where $\sigma'_c = \sigma_c \frac{1}{\rho^2}$ and $\sigma_b' = \sigma_b \frac{1}{\rho^2} $. 
% In particular, we consider three attack methods. Hence, we have three kinds of $\sigma_b$. Let $\sigma_{D2I}$ denote the variance of D2I attack, $\sigma_{iD2D}$ denote the variance of In-D2I attack, $\sigma_{oD2I}$ denote the variance of Out-D2I attack. 
In reality, the variance of clean generations $\sigma_c$ can be a much larger value than that of the backdoor generations. Based on the assumption that $\sigma_c - \sigma_b > \rho^2$, we then have
% \begin{equation}
% % \label{equ:rho}
%     0=\frac{1}{\rho^2}\sigma_{D2I} < \frac{1}{\rho^2}\sigma_{In-D2D} \approx \frac{1}{\rho^2}\sigma_{Out-D2D} < \frac{1}{\rho^2}(\sigma_c-\rho^2) = \frac{1}{\rho^2}\sigma_c - 1.
%     \label{eqn:comparison}
% \end{equation}
% Equation~\ref{eqn:comparison} shows the different variances of the images generated from the fixed diffusion model $f_\theta$ with noise added to the input in terms of clean and backdoor samples. Therefore, we have 
that $\sigma_c' > \sigma_b' + 1$.

% Under Lemma~\ref{lemma:distance}, which states that two points sample from a distribution of larger variance tend to have a larger distance than those from a distribution of lower variance, we have that after noise addition the backdoor inputs tend to generate similar images, while the clean inputs tend to generate diversified images. Take TrojDiff(In-D2D) attack on Cifar10 for example, $N$ is chosen as $32\times32\times3 = 3072$. Based on Lemma~\ref{lemma:distance}, we must have that $\|x_1-x_2\|_2 \geq \|x_3 - x_4\|_2$ under the condition of $\sigma_{In-D2D} < \frac{3071}{3072}\sigma_{c}$, where $x_1, x_2 \sim \mathcal{N}(x_c,\sigma_c \frac{1}{\rho^2} I)$ and $x_3, x_4 \sim \mathcal{N}(x_b,\sigma_{In-D2D} \frac{1}{\rho^2} I)$. However, according to Equation~\ref{eqn:comparison}, this conditional is trivial to achieve, since $\frac{1}{\rho^2} \sigma_c \gg \frac{1}{\rho^2} \sigma_{In-D2D}$. 
This completes the proof.

    \end{proof}
\end{theorem}

\begin{lemma}[Bounds on Expected Length of Gaussian Random Variable~\cite{chandrasekaran2012convex}]
    Given that $x \sim \mathcal{N}(\mu, \sigma I)$, where $x$ is a $N$-dimensional vector. Let $\mathbb{E}(X)$ be the expectation value of the random variable $X$. Then, we have 
    $$\frac{N}{\sqrt{N+1}}\leq \sigma^{-1} \mathbb{E}(\|x\|_2) \leq \sqrt{N}$$
    % , where $\mathbb{E}(X)$ denotes an expectation value of the random variable $X$.
    \label{lemma:norm_bound}
\end{lemma}

% (S1) With at least 99\% probability:  $$\|x_1-x_2\|_2 - \|x_3 - x_4\|_2 \geq \frac{100\sqrt{2}N(\sigma_c - \sigma_b) + 100\sigma_c}{99\sqrt{N+1}} = \mathcal{O}(\sqrt{N}).$$

% (S2) $\mathbb{E}[\|x_1-x_2\|_2 - \|x_3 - x_4\|_2]] \geq R$

\begin{corollary}
    % \label{corollary:large_quantify}
    Under the Theorem~\ref{theorem:dm}, for clean generations $x_1, x_2 \overset{\text{i.i.d.}}{\sim} \mathcal{N}(\mu_c, \sigma_c)$, and backdoor generations $x_3, x_4 \overset{\text{i.i.d.}}{\sim} \mathcal{N}(\mu_b, \sigma_b)$, we have the following statement
     $$
        \mathbb{E}(\|x_1 - x_2\|_2 - \|x_3-x_4\|) \geq \frac{N(\sigma_c - \sigma_b) - \sigma_b}{\sqrt{N+1}} > 0.
    $$

    \begin{proof}

     It is obvious that $x_1 - x_2 \sim \mathcal{N}(0, \sqrt{2}\sigma_c)$ and $x_3 - x_4 \sim \mathcal{N}(0, \sqrt{2}\sigma_b)$. Then,
     According to Lemma~\ref{lemma:norm_bound}, we obtain that
        \begin{equation}
        \begin{split}
            \mathbb{E}(\|x_1 - x_2\|_2 - \|x_3-x_4\|) &= \mathbb{E}(\|x_1 - x_2\|_2) - \mathbb{E}(\|x_3-x_4\|) \\
            &\geq \sqrt{2}( \frac{N\sigma_c}{\sqrt{N+1}} - \sqrt{N}\sigma_b)\\
            &= \sqrt{2}\frac{N\sigma_c - \sqrt{N(N+1)} \sigma_b}{\sqrt{N+1}} \\
            &\geq \sqrt{2}\frac{N\sigma_c - (N+1) \sigma_b}{\sqrt{N+1}} \\
            &=\sqrt{2}\frac{N(\sigma_c -\sigma_b) - \sigma_b}{\sqrt{N+1}}
        \end{split}
        \end{equation}
    Based on the conclusion from the Theorem~\ref{theorem:dm}, we have that $\sigma_c - \sigma_b > 1$. Therefore, we obtain that,
    \begin{equation}
        \mathbb{E}(\|x_1 - x_2\|_2 - \|x_3-x_4\|) > \sqrt{2}\frac{N - \sigma_b}{\sqrt{N+1}} \geq  0,
    \end{equation}
    where the last inequality hold because $\sigma_b = \mathcal{O}(N)$.
    This completes the proof.
    \end{proof}
\end{corollary}

\end{document}